\documentclass{aa}
\usepackage[varg]{txfonts}
\usepackage{color}
\usepackage{amsmath}
\usepackage{mathtools}
\usepackage{float}
\usepackage{natbib}
\usepackage{graphicx}
\usepackage{graphicx}
\usepackage{ulem}
\usepackage{lscape}
\usepackage{xspace}

\usepackage[colorlinks = true,
            linkcolor = blue,
            urlcolor  = blue,
            citecolor = blue,
            anchorcolor = blue]{hyperref}

\usepackage{makecell}

\definecolor{magenta}{rgb}{1.0, 0.0, 1.0}
\definecolor{mygray}{gray}{0.6}

\def\h2o{H$_2$O}

\def\kgm3{kg\,m$^{-3}$}
\def\Jkg3{J\,kg$^{-3}$}

\begin{document} 

    \title{How planets grow by pebble accretion}
   \subtitle{IV: Envelope opacity trends from sedimenting dust and pebbles}
   
   \author{M. G. Brouwers \inst{1}
          \and
          C. W. Ormel \inst{2}
          \and
          A. Bonsor \inst{1}
          \and
          A. Vazan \inst{3, 4}
          }

   \institute{Institute of Astronomy, University of Cambridge, Madingley Road, Cambridge CB3 0HA
   \and Department of Astronomy, Tsinghua University, Haidian DS 100084, Beijing, China
   \and Department of Natural Sciences, Open University of Israel, 4353701 Raanana, Israel
   \and 
   Astrophysics Research Center of the Open university (ARCO), The Open University of Israel, P.O Box 808, Ra’anana, Israel
   \\
              \email{mgb52@cam.ac.uk}
             }

  \abstract
{In the theory of pebble accretion, planets form by the subsequent accretion of solids (micron-sized dust and larger pebbles) and gas. The amount of nebular gas that a planet can bind is limited by its cooling rate, which is set by the opacity of its envelope. Accreting dust and pebbles contribute to the envelope opacity and, thus, influence the outcome of planet formation.}
{Our aim is to model the size evolution and opacity contribution of solids inside planetary envelopes. We then use the resultant opacity relations to study emergent trends in planet formation.}
{We design a model for the opacity of solids in planetary envelopes that accounts for the growth, fragmentation and erosion of pebbles during their sedimentation. It also includes a separate dust component, which can be both replenished and swept up by encounters with pebbles, depending on the relative velocities. We formulate analytical expressions for the opacity of pebbles and dust and map out their trends as a function of depth, planet mass, distance and accretion rate.}
{The accretion of pebbles rather than planetesimals can produce fully convective envelopes, but only in lower-mass planets that reside in the outer disk or in those that are accreting pebbles at a high rate. In these conditions, pebble sizes are limited by fragmentation and erosion, allowing them to pile up in the envelope. At higher planetary masses or reduced accretion rates, a different regime applies where the sizes of sedimenting pebbles are only limited by their rate of growth. The opacity in this growth-limited regime is much lower, steeply declines with depth and planet mass but is invariant with the pebble mass flux. Our results imply that the opacity of a forming planet's envelope can not be approximated by a value that is constant with either depth or planet mass. When applied to the Solar System, we argue that Uranus and Neptune could not have maintained a sufficiently high opacity to avoid runaway gas accretion unless they both experienced sufficiently rapid accretion of solids and formed late.}
{}
   \keywords{Planetary systems -- Planets and satellites: composition -- Planets and satellites: formation -- Planets and satellites: physical evolution -- Planet-disk interactions}
   
   \maketitle
   
\section{Introduction}\label{sect:introduction}

Pebble accretion is a version of core accretion where planets grow primarily by intercepting a stream of sub-cm sized particles \citep{Ormel2010, Lambrechts2012}. Previous works have mainly considered its dynamical aspects and have demonstrated that it can provide a rapid channel for growth \citep{Morbidelli2012, Chambers2014}, especially when pebbles are settled in the mid-plane and accretion proceeds in a 2D manner \citep{Liu2018, Ormel2018}. A key feature of pebble accretion is that smaller particles are naturally prevented from reaching the planet once it has grown sufficiently large to perturb the surrounding disk \citep{Morbidelli2015, Bitsch2018, Eriksson2020}. This has been suggested as a way of discerning gas and ice giant formation \citep{Lambrechts2014}, and of limiting the growth of super-Earths around low-mass stars \citep{Liu2019, Chen2020}. Its drag-based mode of capture also introduces an asymmetry that might explain the structural prograde rotation of smaller solar system objects \citep{Johansen2010, Visser2020}. \let\thefootnote\relax\footnotetext{Corresponding author: Chris Ormel (chrisormel@tsinghua.edu.cn)}

Besides these global effects, the accretion of smaller particles also has important consequences for a planet's internal structure. So far, most inquiries have focused on compositional changes, as the rapid vaporization of pebbles \citep{Love1991, Mcauliffe2006} naturally leads to the deposition of significant amounts of vapor, which can be stored in the deep interior \citep{Iaroslavitz2007, Lozovsky2017}. This naturally limits the size of central cores \citep{Alibert2017, Brouwers2018} and results in a dense, polluted interior \citep{Iaroslavitz2007, Venturini2016, Lozovsky2017, Bodenheimer2018}. The traditional rigid core-envelope structure of these planets is replaced by the natural emergence of a compositional gradient \citep{Valletta2020, Ormel2021, Vazan2020}. This seems to be in line with Juno's measurements of Jupiter's gravitational moments of inertia, which infer such a dilute core structure \citep{Wahl2017, Debras2019,Debras2021}. 

Before pebbles get to these inner regions, however, they must first sediment through the tenuous outer layers where their combined surface area can contribute to the opacity. This influence of pebbles on the outer envelope is still largely unexplored, but has important thermodynamic consequences. The more opaque an envelope, the less heat is able to escape it and the less gas it can gravitationally bind. By regulating the pace of cooling, the opacity is a key variable that directly affects the outcome of planet formation. The quantitative importance of the opacity is often understated in current formation models, where a lack of physically motivated values is still a serious issue. It is common practice to adopt conveniently high ISM-like opacities or values that are arbitrarily scaled down, with important consequences for the outcome of these models. One of the reasons that envelope opacity values are currently poorly constrained, is that previous grain growth models focused specifically on the formation of Jupiter at 5 AU. In the most detailed of these, \citet{Podolak2003, Movshovitz2008, Movshovitz2010} calculated grain growth with numerical models that solve the Smoluchowski equation under the assumption that grains stick when they collide. They found that grain growth effectively reduces the dust opacity as a function of depth, with the opacity declining by around three orders of magnitude from the outer layers to the inner radiative-convective boundary (RCB). Simpler analytical \citep{Mordasini2014b} and numerical \citep{Ormel2014} models with a single characteristic grain size at a given height have been able to replicate this main result. However, because proto-planetary disks are thermodynamically very different across distances and masses, it is not clear that these results can be applied generally to planets of varying sizes throughout the disk.

In addition, whether a planet is predominantly accreting sub-cm pebbles or 100-km planetesimals is clearly important for the abundance of grains in the outer envelope. The assumption of grain growth models that planetesimals deposit significant large amounts of small grains in the upper layers is in conflict with impact simulations, which predict that most of their mass is released close to the planet's central cores, far below the RCB \citep{Brouwers2018, Valletta2019, Valletta2020}. In contrast, \citet{AliDib2020, Johansen2020} have recently shown that pebbles are very susceptible to erosion by small dust grains when they enter planetary envelopes, especially when they are accelerated further by convective cells. It was also pointed out by \citet{Ormel2021} that even without any size evolution, mm-sized pebbles can contribute a significant opacity to planetary envelopes if they accrete at a sufficiently high rate.

The total envelope opacity is a sum of the contributions of solids and gas ($\kappa = \kappa_\mathrm{s} + \kappa_\mathrm{gas}$). In this study, we formulate a model for the contribution of solids, which often dominates over the gas during accretion. In Sect. \ref{sect:basic_model}, we develop a physical opacity model that accounts for the main processes that influence the evolution of solids in planetary envelopes. We consider a variation to the single-size approximation where we model the populations of both small dust ($\kappa_\mathrm{d}$) and larger pebbles ($\kappa_\mathrm{peb}$). The characteristic size of the pebbles in our model is either determined by their growth from sticking collisions (coalescence) at low velocities or limited by fragmentation and erosion at high velocities, depending on the local thermodynamic conditions. When pebbles travel at speeds below the erosion velocity, they sweep up dust grains without experiencing mass loss \citep{Krijt2015, Schrapler2018}. We assume a constant dust size and calculate its steady state abundance between dust sweep-up and production, which can occur as a by-product from collisions between pebbles or by erosion. In Sect. \ref{sect:analytical_expressions}, we then formulate simple, physically motivated analytical expressions for the opacity from solids, similar to what was done by \citet{Mordasini2014b}. We apply our opacity model across a wide parameter space to explore and map out the resulting opacity as a function of distance, planet mass and accretion rate in Sect. \ref{sect:opacity_trends}. We then present the implications of these findings to the formation of gas and ice giants in Sect. \ref{sect:implications}. We discuss our model in relation to contemporary works, list potential model improvements and contrast the scenarios of pebble and planetesimal accretion in Sect. \ref{sect:discussion}. Finally, we conclude our work in Sect. \ref{sect:conclusions}. In appendix \ref{appendix_A}, we derive an expression for the non-isothermal critical metal mass that we use to study the onset of runaway growth. In appendix \ref{sect:appendix_B}, we motivate our constant dust size by comparing how far dust grains can travel before they collide. In the final appendix (\ref{sect:appendix_C}), we vary the limiting velocity, the parameter in our model that represents the onset of fragmentation as well as erosion by micron-sized grains.

\section{Model description}\label{sect:basic_model}
\begin{figure}[t!] 
\centering
\includegraphics[width=\hsize]{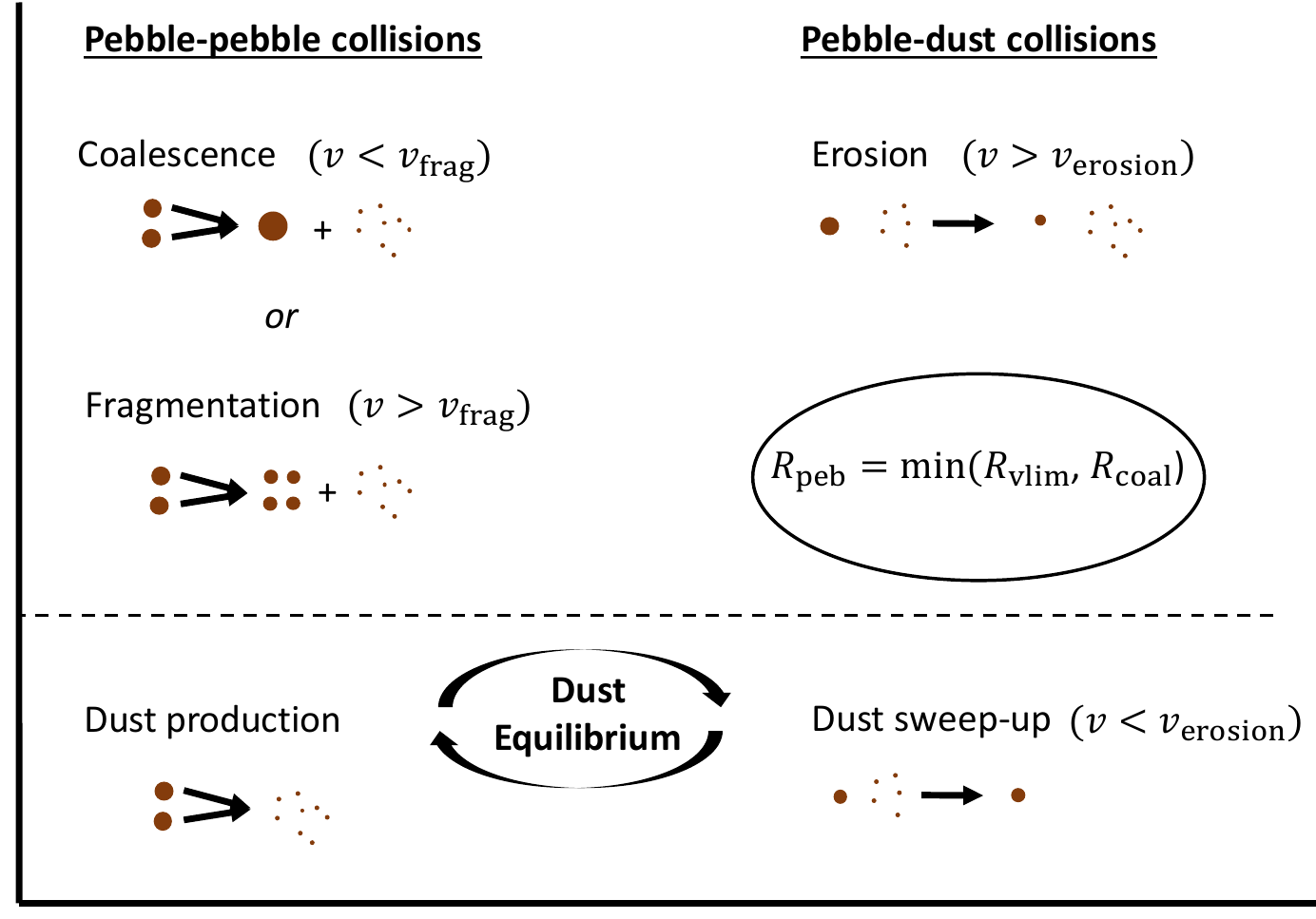} 
\caption{Qualitative sketch of our two-population opacity model. Depending on the velocities of sedimenting pebbles, they can experience growth (coalescence), fragmentation and erosion. In the growth-limited regime, the pebble size in the interior is regulated by the collision and sedimentation timescales ($R_\mathrm{peb} = R_\mathrm{coal}$). In the velocity-limited regime, fragmentation and/or erosion restrict the pebble size below the growth potential ($R_\mathrm{peb} = R_\mathrm{vlim} < R_\mathrm{coal}$). In our model, dust grains of constant size are produced in collisions between pebbles or in high-velocity pebble-dust encounters (erosion), while they are lost by sticking encounters with slow-moving pebbles (dust sweep-up), leading to a local steady state in their abundance. \label{fig:sketch}} 
\end{figure}

\subsection{Two-population approach}
Growing proto-planets are supplied with a flux of solid particles that enter their envelopes by two distinct mechanisms. The first is the drag-assisted capture of sub-cm pebbles that drift towards the central star \citep{Ormel2010, Lambrechts2012}, supplemented by the accretion of larger, km-sized planetesimals \citep{Alibert2018, Guilera2020}. The second source is the population of small, micron-sized dust grains that are coupled to the nebular gas and that enter the planet's envelope when it is massive enough to bind one.

In two recent works, \citet{AliDib2020, Johansen2020} have shown that the interaction between dust and pebbles is crucial in evaluating their relative abundances, as high-velocity collisions between the two can significantly erode the incoming pebbles and add to the dust population \citep{Krijt2015,Schrapler2018}. This erosion is a runaway process and is only halted at the pebble size where collisions switch to sticking and the dust abundance begins to drop. In our model, we incorporate both dust and pebbles in a simple manner where each is represented with a single characteristic size. 
The pebble size is determined locally by collisions within the population (growth or fragmentation), as well as collisions with the dust (sweep-up and erosion). The size of individual dust particles is a parameter in our model, which we keep constant as motivated by the ongoing production and sweep-up of these smaller grains in the envelopes (see Appendix \ref{sect:appendix_B}). Fig. \ref{fig:sketch} provides a qualitative overview of our model, which we work out in the next subsections.

\subsection{Sedimentation of solids}\label{sect:sedimentation}
Particles referred to as dust and pebbles are small enough to be effectively slowed down by gas drag and sink at speeds close to their terminal velocities ($v_\mathrm{fall}$) relative to the local medium. We adopt a simple, continuous two-regime approximation to the drag force, where the transition from free molecular (Epstein) to continuum flow (Stokes) is set at the canonical value $l_\mathrm{mfp}/R_\mathrm{s} = 4/9$ and the terminal velocity is given by \citep{Weidenschilling1977a}:
\begin{equation}\label{eq:vfall}
    v_\mathrm{fall} = \frac{gR_\mathrm{s}\rho_\mathrm{\bullet}}{\rho_\mathrm{g} v_\mathrm{th}} \; \mathrm{max\left(\frac{4R_\mathrm{s}}{9l_\mathrm{mfp}} , 1\right)},
\end{equation}
where $g$ and $\rho_\mathrm{g}$ are the local gravitational acceleration and gas density at a distance from the planet's center $r$ and the subscript $s$ refers to either population of solids (pebbles with size $R_\mathrm{peb}$ and dust with size $R_\mathrm{d}$), which share the same material density $\rho_\mathrm{\bullet}$. We take the standard ideal gas expressions for the thermal velocity $v_\mathrm{th}$ and the mean free path of molecules $l_\mathrm{mfp}$. The second component to the sedimentation velocity of solids is given by the downward flow of gas. Because we are mostly interested in the planet's upper layers that only contain a fraction of the envelope mass, we can approximate this gas velocity ($v_\mathrm{gas}$) from mass conservation of the gas accretion rate ($\dot{M}_\mathrm{xy}$) as:
\begin{equation}\label{eq:vgas}
    v_\mathrm{gas} = \frac{\dot{M}_\mathrm{xy}}{4\pi r^2 \rho_\mathrm{g}}.
\end{equation}
\begin{table}[t]
\caption{Descriptions and values of the default model parameters. The disk conditions are adopted from the Minimum Mass Solar Nebula and scale as $T_\mathrm{disk} \propto d^{-1/2}, \; \rho_\mathrm{disk} \propto d^{-11/4}$ \citep{Weidenschilling1977b, Hayashi1981}.}
\label{table_parameters}
\centering
\addtolength{\tabcolsep}{-0.9 mm}
\begin{tabular}{l l l}
\Xhline{2\arrayrulewidth}
\Xhline{2\arrayrulewidth}
Parameter & Description & Value \\ 
\Xhline{2\arrayrulewidth}   
$M_\mathrm{p}$ & Total planet mass  & $5 \; \mathrm{M_\oplus}$ \\     
$M_\mathrm{c}$ & Central core mass & $2 \;  \mathrm{M_\oplus}$ \\
$\rho_\mathrm{c}$ & Central core density & $3.2 \;\mathrm{g /cm^3}$ \\
$\rho_\mathrm{\bullet}$ & Dust and pebble density & $3.2 \; \mathrm{g /cm^3}$ \\
$\dot{M}_\mathrm{peb}$ & Pebble accretion rate  & $10^{-6} \; \mathrm{M_\oplus/yr}$ \\ 
$\dot{M}_\mathrm{xy}$ & Gas accretion rate  & $10^{-7} \; \mathrm{M_\oplus/yr}$ \\ 
$R_\mathrm{d}$ & Dust monomer radius  & $1 \; \mathrm{\mu m}$ \\   
$F$ & Dust replenishment constant  & 0.1 \\   
$v_\mathrm{frag}$ & Fragmentation velocity  & 0.8 m/s\\  
$T_\mathrm{vap}$ & Sublimation temperature  & 2500 K \\  
$d$             & Orbital distance          & 5 AU \\ 
$M_\star$      & Mass of the central star & $1 \; \mathrm{M_\odot}$  \\
$\rho_\mathrm{disk, 5AU}$     & Disk density at 5 AU   & $5 \times \, 10^{-11}$ $\mathrm{g /cm^3}$                \\
$T_\mathrm{disk,5AU}$     & Disk temperature  at 5 AU  & 150 K               \\
$\mu_\mathrm{xy}   $      &  Molecular weight nebular gas  & 2.34 $\; \mathrm{m_H}$            \\ 
$\nabla_\mathrm{ad}   $      &  Adiabatic temperature gradient   & 0.31                      \\ 
\Xhline{2\arrayrulewidth}                               
\end{tabular}
\end{table}
When the solids are small enough, as is the case for the micron-sized dust particles in our model, the local downward flow of gas dominates their total sedimentation velocity. This was described as the advection regime in the work by \citet{Mordasini2014b}. For larger particles, the free-fall term instead dominates and generally, the total sedimentation velocity relative planet's core is a sum of the two:
\begin{equation}\label{eq:vsed}
    v_\mathrm{sed} = v_\mathrm{fall} + v_\mathrm{gas}.
\end{equation} 
We consider quasi-static envelopes, where the grain transport timescale is assumed to be short relative to the timescale on which the envelope's thermodynamic conditions change. This is a good assumption in the outer layers, provided that the total envelope mass is dominated by the planet's polluted interior. If the solids additionally enter in a spherically symmetric manner, their volume density is given by:
\begin{equation}
   \rho_\mathrm{s} = \frac{\dot{M}_\mathrm{s}}{4\pi r^2 v_\mathrm{sed}},
\end{equation}
which represents radial mass conservation. We do not include large-scale convective motions \citep{Popovas2018, Popovas2019} in our model, which could locally transport clumps of solids and alter their collision velocities \citep{Ormel2007}.

\subsection{Dust-pebble collisions}
\subsubsection{Erosion at high velocities}\label{sect:erosion}
\begin{figure}[t!] 
\centering
\includegraphics[width=\hsize]{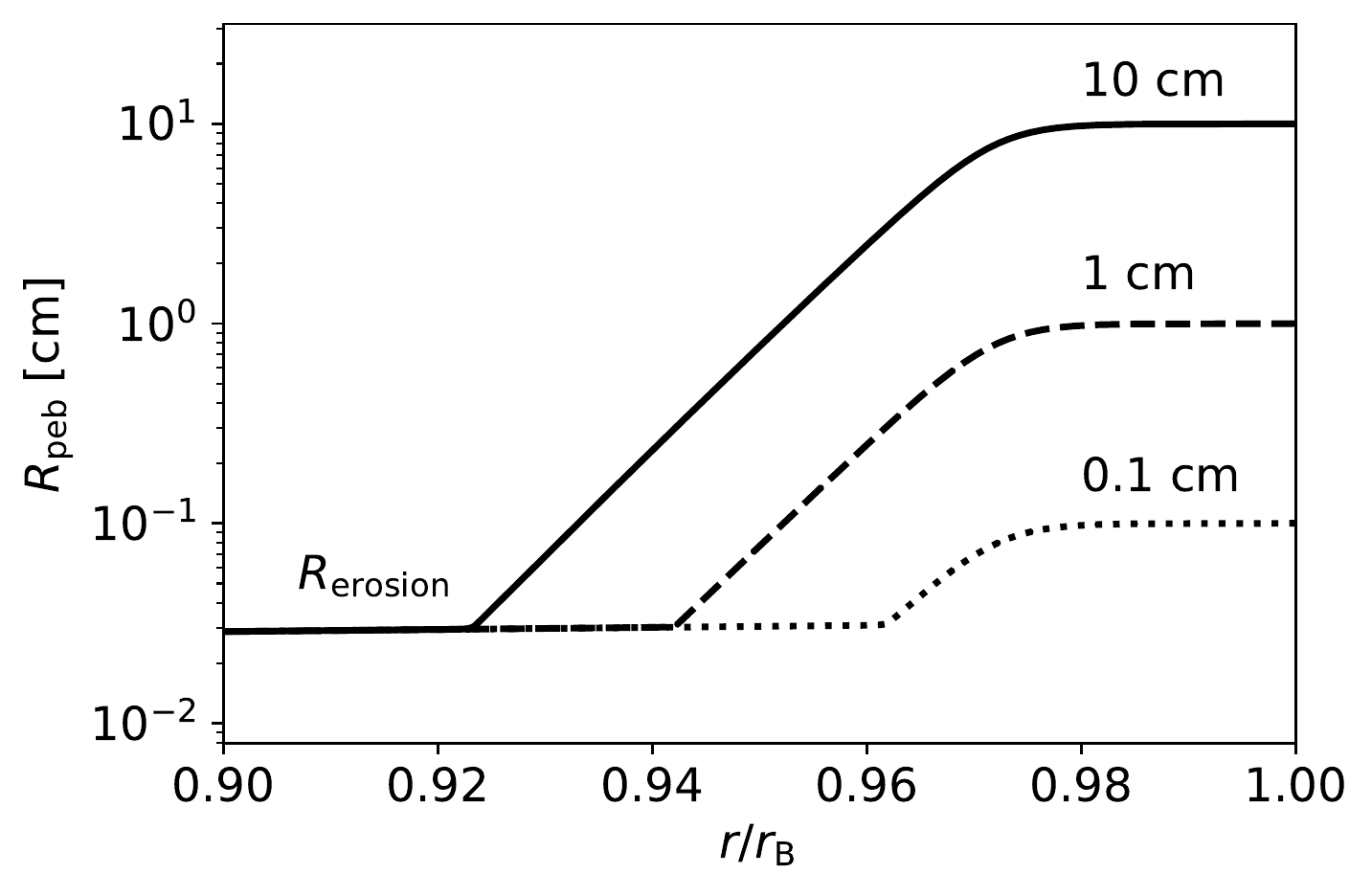} 
\caption{Erosive mass loss of pebbles that enter envelope of a growing planet at its Bondi radius ($r_\mathrm{B}$). The parameters of the planet are shown in Table \ref{table_parameters}. The three lines correspond to initial pebble entry sizes of 0.1 cm (dotted), 1 cm (dashed) and 10 cm (solid). Due to the positive feedback of erosive mass-loss on the dust abundance, the pebbles rapidly converge to the size $R_\mathrm{erosion}$ (Eq. \ref{eq:R_erosion}) where collisions with dust grains switch from causing mass loss by erosion to sticking (sweep-up).\\ \label{fig:erosion}} 
\end{figure}

The sizes of pebbles that enter planetary envelopes are determined by their growth, fragmentation and radial drift in the proto-planetary disk \citep[i.e.][]{Liu2020, Drazkowska2021}. The observational evidence points towards typical sizes between $100 \; \mathrm{\mu m} - 1 \; \mathrm{cm}$ \citep[i.e.][]{Birnstiel2010, Kataoka2016, Hull2018, Carrasco2019, Ohashi2020, Tazzari2020a}, which generally decrease with distance to the central star \citep{Tazzari2020b}. During their capture by the planet, pebbles are subject to additional gravitational acceleration and begin to collide with dust at increasing velocities. It was pointed out by \citet{AliDib2020, Johansen2020} that if this velocity crosses the threshold for erosion, the dust particles begin to chip off pebble material. This process was measured by \citet{Schrapler2018} to begin at:
\begin{equation}\label{eq:verosion}
    v_\mathrm{erosion} = 2.4 \; \mathrm{m/s} \left(\frac{R_\mathrm{d}}{1 \; \mathrm{\mu m}}\right)^\frac{1}{1.62},
\end{equation}
with an erosive mass loss per collision $\Delta m_\mathrm{erosion}$ of:
\begin{equation}
    \frac{\Delta m_\mathrm{erosion}}{m_\mathrm{d}} = 4.3 \left(\frac{v}{10 \; \mathrm{m/s}}\right) \left(\frac{R_\mathrm{d}}{1 \; \mathrm{\mu m}}\right)^{-0.62},
\end{equation}
which requires a mass ratio below $\sim 10^{-2}$ \citep{Krijt2015}. The proportionality of the erosion efficiency to the collision velocity was also found in numerical investigations \citep{Seizinger2013, Planes2017}. The fragments that are produced are generally similar or smaller in size than the projectiles, which in our model are dust grains with mass $m_\mathrm{d}$. Because the produced fragments add to the dust population, the erosion of pebbles is a runaway process up to a characteristic size where they slow down sufficiently by their increased susceptibility to gas drag. In our model with linear downward sedimentation, this is set by the condition that $v_\mathrm{peb, fall}=v_\mathrm{erosion}$ (assuming $v_\mathrm{fall, d} \ll v_\mathrm{fall, peb}$):
\begin{equation}\label{eq:R_erosion}
    R_\mathrm{erosion} = \left\{\begin{array}{l}
    \displaystyle\frac{\rho_\mathrm{g} v_\mathrm{th} v_\mathrm{erosion}}{g\rho_\mathrm{\bullet}}
    \mathrm{\; (\mathrm{Epstein})}
    \\
    \\
   \displaystyle\frac{3}{2}{\left(\frac{\rho_\mathrm{g} v_\mathrm{th} l_\mathrm{mfp} v_\mathrm{erosion}}{g\rho_\mathrm{\bullet}}\right)}^\frac{1}{2}
    \mathrm{\; (\mathrm{Stokes})}.
    \end{array}\right.
\end{equation}
In order to estimate how rapidly pebbles erode, we formulate the radial equations of mass transfer and size evolution. The typical distance that pebbles travel before they encounter a dust particle is given by (assuming $R_\mathrm{d} \ll R_\mathrm{peb} \; \mathrm{and} \; v_\mathrm{fall, d} \ll v_\mathrm{fall, peb}$):
\begin{equation}\label{eq:l_sweep}
    l_\mathrm{sweep,peb} = \frac{m_\mathrm{d}}{\rho_\mathrm{d}\sigma_\mathrm{peb}} \frac{v_\mathrm{sed, peb}}{v_\mathrm{fall,peb}},
\end{equation}
where $\sigma_\mathrm{peb} = \pi R_\mathrm{peb}^2$ is the pebble cross section. The equations for mass transfer and shrinkage follow as:
\begin{subequations}\label{eq:erosion_loss}
\begin{align}
    \frac{d\dot{M}_\mathrm{peb}}{dr}\bigg|_\mathrm{erosion}
    &= \frac{\dot{M}_\mathrm{peb}}{l_\mathrm{sweep,peb}} \frac{\Delta m_\mathrm{erosion}}{m_\mathrm{peb}}, \label{eq:erosion_1} \\
    \frac{d\dot{M}_\mathrm{d}}{dr}\bigg|_\mathrm{erosion} &= -\frac{d\dot{M}_\mathrm{peb}}{dr}, \\
    \frac{d m_\mathrm{peb}}{dr}\bigg|_\mathrm{erosion} &= \frac{\Delta m_\mathrm{erosion}}{l_\mathrm{sweep,peb}}, \label{eq:erosion_2}
\end{align}
\end{subequations}
where $\dot{M}_\mathrm{peb}$ is the pebble mass flux. In this simple model, the increase in dust abundance from pebble erosion is essentially exponential and the pebbles shrink very rapidly, even without assuming any additional acceleration from convective eddies. As an example, we integrate Eqs. \ref{eq:erosion_1}-\ref{eq:erosion_2} from the Bondi radius inwards for a nominal planet of 5 $\mathrm{M_\oplus}$ at 5 AU with a dust disk-to-gas ratio of $10^{-3}$ (see Table \ref{table_parameters} for all default parameters). The resulting size evolution curves are shown in Fig. \ref{fig:erosion} for three different pebble entry sizes. In all the runs, the pebbles quickly erode down to the size where their velocity decreases below the threshold for erosion and they are safe. In this manner, erosion essentially cancels out the upper range of the initial pebble size distribution to more uniform pebble entry sizes, limited by their erosion at the Bondi radius.

\subsubsection{Sweep-up at low velocities}
If the pebbles are reduced in size below $R_\mathrm{erosion}$, the relative velocities between the dust and the pebbles drop below $v_\mathrm{erosion}$ and dust can stick to the pebbles without dislodging additional material. In this sweep-up regime, Eqs. \ref{eq:erosion_1} - \ref{eq:erosion_2} still apply with $\Delta m = -m_\mathrm{d}$:
\begin{equation}
    \frac{d\dot{M}_\mathrm{d}}{dr}\bigg|_\mathrm{sweep-up}
    = \frac{\dot{M}_\mathrm{peb}}{l_\mathrm{sweep,peb}} \frac{ m_\mathrm{d}}{m_\mathrm{peb}}, \label{eq:dust_sweepup} 
\end{equation}
Now, the presence of pebbles acts to reduce the dust abundance, rather than increase it. In Sect. \ref{sect:equilibrium}, we will discuss the scenario where the production of dust by erosion and pebble-pebble collisions is in equilibrium with the sweep-up of dust, providing a steady state dust abundance. 

\subsection{Pebble-pebble collisions}
When considering straight downward sedimentation, collisions between two pebbles can occur either through Brownian motion or by differential settling (coalescence). The timescale of the latter process decreases with increasing particle size, such that coalescence easily dominates their mutual collision rate when the pebbles are larger than several microns \citep{Boss1998, Nayakshin2010, Mordasini2014b, Ormel2014}. As identical particles settle with the same speed, coalescence within a population formally arises from the parameterized width in the size distribution. In our model, we follow the works by \citet{Rossow1978,Mordasini2014b} and set the typical mass ratio of collisions to 0.5. Although a simplification, \citet{Krijt2016, Sato2016} show that this is a good approximation to the average mass ratio found in coagulation simulations that treat a complete size distribution. When combined with Eq. \ref{eq:vsed}, it leads to collisional velocities in the range of $0.21<v_\mathrm{col}/v_\mathrm{fall}<0.37$ depending on the drag regime. In our model, we neglect this order unity difference in favor of a continuous collision velocity across drag regimes and approximate it as $x_\mathrm{R} \equiv v_\mathrm{col}/v_\mathrm{fall} \simeq 1/3$.

\subsubsection{Growth by coalescence at low velocities}\label{sect:growth}
When the collision velocities are sufficiently low, contact between pebbles leads to sticking and growth \citep{Dominik1997}. To work out the rate of this growth, the typical travel distance between pebble collisions ($l_\mathrm{col,peb}$) can be related to their local size, abundance and velocity as:
\begin{equation}\label{eq:l_col}
    l_\mathrm{col,peb} = \frac{m_\mathrm{peb}}{ \rho_\mathrm{peb} \sigma_\mathrm{peb}} \frac{v_\mathrm{sed, peb}}{v_\mathrm{col,peb}}.
\end{equation}
Eq. \ref{eq:l_col} scales positively with pebble size, which indicates that smaller particles collide more often. By equating $l_\mathrm{col, peb}$ to the scale height $H=k_\mathrm{b}T_\mathrm{g}/(\mu_\mathrm{g}g)$, which represents the characteristic length scale in an envelope, it is possible to approximate the maximum sizes ($R_\mathrm{coal}$) to which solids can grow as they sediment:
\begin{equation}\label{eq:R_growth}
    \displaystyle R_\mathrm{coal} = \left\{\begin{array}{l}
    \displaystyle{\left(\frac{3 x_\mathrm{R} H \dot{M}_\mathrm{peb} v_\mathrm{th}\rho_\mathrm{g}}{16 \pi G M_\mathrm{p} \rho_\mathrm{\bullet}^2}\right)}^\frac{1}{2}
    \mathrm{\; (\mathrm{Epstein})}
    \\
    \\
    \displaystyle{\frac{3}{4}\left(\frac{x_\mathrm{R} H \dot{M}_\mathrm{peb} v_\mathrm{th} l_\mathrm{mfp} \rho_\mathrm{g}}{\pi G M_\mathrm{p} \rho_\mathrm{\bullet}^2}\right)}^\frac{1}{3}
    \mathrm{\; (\mathrm{Stokes})},
    \end{array}\right.
\end{equation}
where we assumed that their total velocities are dominated by the terminal component rather than by inward gas flow. In the regime where growth by coalescence provides the limit to their size, the pebbles are typically large enough ($R_\mathrm{peb} \gtrsim 100 \; \mathrm{\mu m}$) that this is indeed a good assumption. The scaling of Eq. \ref{eq:R_growth} with the pebble accretion rate shows that a greater mass flux of pebbles leads to more collisions and faster growth. Its dependence on the planet's mass is more complicated. More massive planets have a stronger gravitational pull, which increases the distance at which they can bind gas and, therefore, extends the envelope ($r_\mathrm{B} \propto M_\mathrm{p}$). This lowers the gravitational acceleration at the planet's outer boundary as a function of mass ($g \propto M_\mathrm{p}/r_\mathrm{B}^2 \propto M_\mathrm{p}^{-1}$). Taken together, the collision distance scales as $l_\mathrm{col,peb} \propto M_\mathrm{p}$, which balances with the increasing scale height $H \propto M_\mathrm{p}$ and leads to a constant value of $R_\mathrm{coal}$ across planetary masses.

\subsubsection{Collisional fragmentation at high velocities}\label{sect:fragmentation}
\begin{figure}[t!] 
\centering
\includegraphics[width=\hsize]{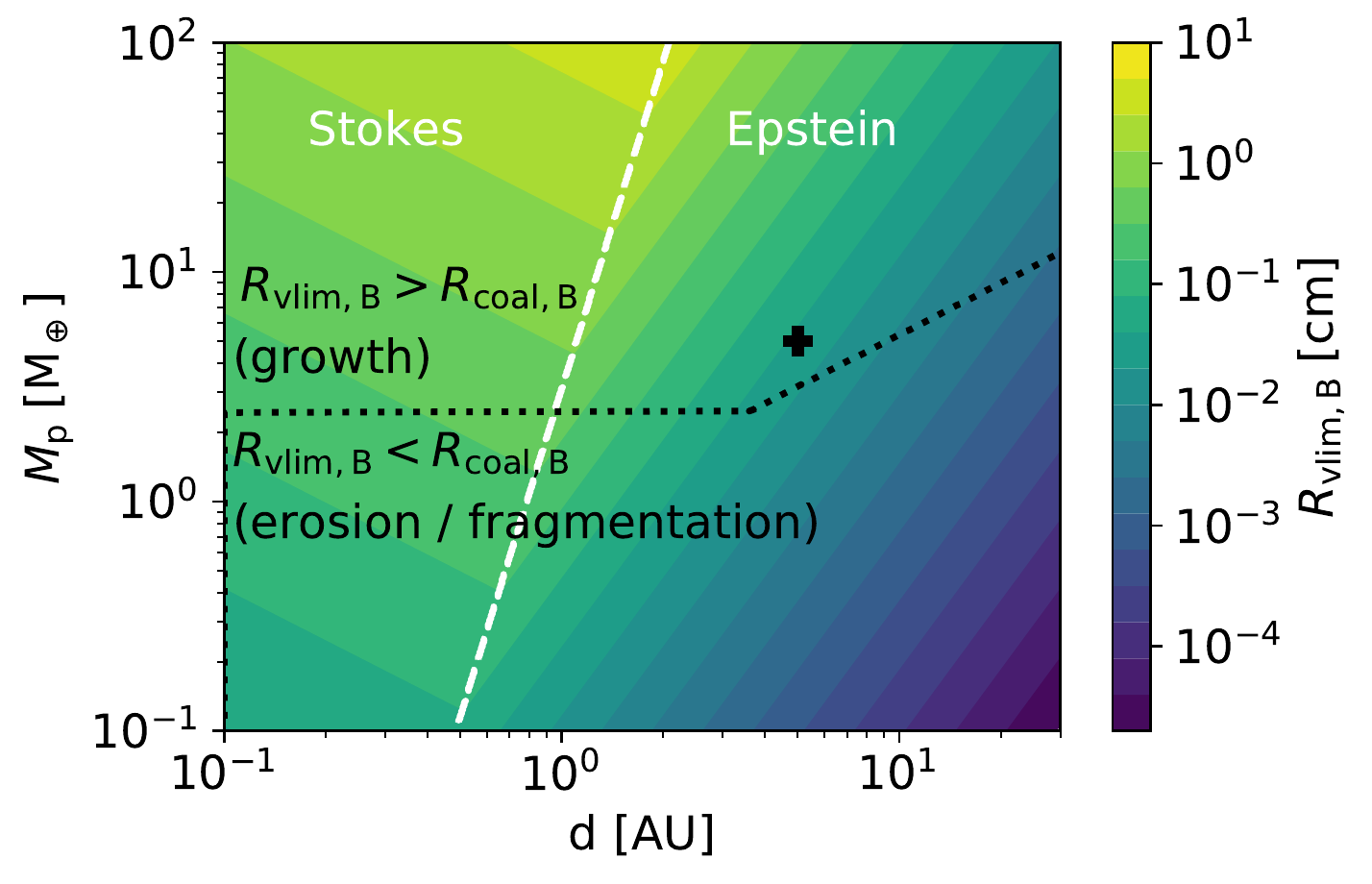}
\caption{Characteristic pebble size limits at the Bondi radius, plotted with the parameters of Table \ref{table_parameters}. The colors indicate the pebble size where $v_\mathrm{fall,peb} = v_\mathrm{lim} = 2.4 \; \mathrm{m/s}$, the common limit set by erosion with micron-sized grains and fragmentation. The white dashed line separates the Stokes and Epstein regimes while the dotted line indicates the regions where either growth rate or erosion/fragmentation limits the local pebble size. The plus sign marks the default planet mass ($5 \; \mathrm{M_\oplus}$) and distance (5 AU).
\label{fig:erosion_fragmentation}} 
\end{figure}

At higher velocities, collisions between pebbles lead to bouncing or even fragmentation. Collision experiments typically find that silicate particles begin to fragment if their collision velocity is larger than 1 m/s \citep{Wurm2005, Schafer2007, Guttler2010}, and up to 10 m/s for water ice grains \citep{Gundlach2015, Musiolik2019}. This fragmentation criterion can be combined with the velocity for the onset of erosion (see Eq. \ref{eq:verosion}) to yield a limit on the terminal velocities of pebbles:
\begin{equation}\label{eq:vlim}
    v_\mathrm{lim} = \mathrm{min}\left(v_\mathrm{erosion}, v_\mathrm{frag}/x_\mathrm{R}\right).
\end{equation}
Depending on the dust size and the pebble's material properties, either fragmentation or erosion can be the limiting factor to pebble growth. For simplicity, we take a default dust size of 1 $\mathrm{\mu m}$ and a fragmentation velocity of $0.8$ m/s, for which the two velocity limits on the pebble's terminal velocity are both equal to $v_\mathrm{fall,peb}=v_\mathrm{lim}=2.4$ m/s. In appendix \ref{sect:appendix_C}, we vary the limiting velocity within a wider range. We refer to the velocity-limited pebble size as $R_\mathrm{vlim}$ and it can be found from Eq. \ref{eq:vfall} as:
\begin{equation}\label{eq:R_crit}
    R_\mathrm{vlim} =\left\{\begin{array}{l}
    \displaystyle \frac{\rho_\mathrm{g} v_\mathrm{th} v_\mathrm{lim}}{g\rho_\mathrm{\bullet}}
    \mathrm{\; (\mathrm{Epstein})}
    \\
    \\
   \displaystyle \frac{3}{2}{\left(\frac{\rho_\mathrm{g} v_\mathrm{th} l_\mathrm{mfp} v_\mathrm{lim}}{g\rho_\mathrm{\bullet}}\right)}^\frac{1}{2}
    \mathrm{\; (\mathrm{Stokes})}.
    \end{array}\right.
\end{equation}
The dependence of $R_\mathrm{vlim}$ at the Bondi radius is visualized in Fig. \ref{fig:erosion_fragmentation}, with comparison to the limit set by the growth rate. In contrast to $R_\mathrm{coal}$, the value of $R_\mathrm{vlim}$ scales positively with the planet's mass. Hence, the sizes of pebbles in small envelopes are typically velocity-limited by erosion or fragmentation, whereas pebbles in more massive envelopes are typically only limited by their rate of growth. Close to the central star, the disk is relatively dense and pebbles of the same size sediment more slowly. Consequently, pebbles erode or fragment down to the smallest sizes when they accrete onto low-mass planets that reside in the outer disk.

Besides limiting the sizes to which pebbles can grow, collisions between pebbles can also potentially result in production of small dust as collisional by-products. In our two-component model, we include this with a fractional dust production efficiency $F$, which can theoretically be between 0 (no dust production) and 1 (all growth beyond $R_\mathrm{vlim}$ is turned into dust). Physically, F is a measure for the number of pebble-pebble collisions needed to completely grind down a pebble. The rate at which this dust is produced is then given by the pebble collision rate as:
\begin{subequations}\label{eq:fragmentation_loss}
\begin{align}
    \frac{d\dot{M}_\mathrm{d}}{dr}\bigg|_\mathrm{frag}
    &= -F \frac{\dot{M}_\mathrm{peb}}{l_\mathrm{col,peb}}, \\
    \frac{d m_\mathrm{peb}}{dr}\bigg|_\mathrm{frag} &= F \frac{m_\mathrm{peb}}{l_\mathrm{col,peb}}.
\end{align}
\end{subequations}
We will discuss reasonable values for the parameter $F$ in Sect. \ref{sect:opac_dust}.

\section{Analytical opacity expressions}\label{sect:analytical_expressions}
In this section, we formulate analytical expressions for the opacity contributions from pebbles and dust. For the pebbles, these follow from the previously identified pebble size limits in different regimes ($R_\mathrm{coal}, R_\mathrm{vlim}$). The basic expression for the Rosseland mean opacity from solids ($\kappa_\mathrm{s}$) is:
\begin{equation}\label{eq:f_kappa}
    \kappa_\mathrm{s} = \frac{3Q_\mathrm{eff} \rho_\mathrm{s}}{4\rho_\mathrm{\bullet} R_\mathrm{s} \rho_\mathrm{g}},
\end{equation}
where the extinction efficiency $Q_\mathrm{eff}$ can be approximated as $Q_\mathrm{eff} \simeq \mathrm{min}\big(0.6\pi R_\mathrm{s}/\lambda_\mathrm{peak}, 2\big)$, with the peak wavelength of the emitted photons by the local gas $\lambda_\mathrm{peak} (\mathrm{cm})= 0.290/T_\mathrm{g}$ (Wien's law). Laboratory experiments provide a more detailed temperature scaling of $Q_\mathrm{eff}$, as well as a factor $\sim$ 2 between different species \citep{Movshovitz2008, Bitsch2021}. We do not include these additional details here but do account for the most important opacity difference between species, which is that solid particles are only present in layers of planetary envelopes that are sufficiently cool for the solids to escape sublimation. In the case of silicates, which we consider as the default composition of pebbles and dust, this threshold is positioned deep inside planetary envelopes around $\sim$ 2500 K.

\subsection{Pebble opacity expression}
The first opacity regime is the growth-limited regime, which applies when pebbles are traveling relatively slowly through planetary envelopes and experience sticking collisions, rather than erosion or fragmentation. In this case, their size can be approximated by $R_\mathrm{coal}$ (Eq. \ref{eq:R_growth}) and their opacity contribution is independent of their accretion rate. It follows the same equation in both drag regimes:
\begin{equation}\label{eq:opac_coal}
    \kappa_\mathrm{coal} = \frac{Q_\mathrm{eff}}{x_\mathrm{R} H \rho_\mathrm{g}},
\end{equation}
which was the main finding of \citet{Mordasini2014b}. As indicated by Fig. \ref{fig:erosion_fragmentation} at the Bondi radius, this growth-limited regime is mainly applicable in the envelopes of more massive planets, whose larger envelopes allow pebbles to sediment more slowly. The required planetary size to enter this regime at the same accretion rate is an increasing function of distance from the central star, as pebbles in the tenuous outer disk sediment faster than those closer in. When the conditions are such that the terminal velocity of pebbles exceeds $v_\mathrm{lim}$, the size to which solids can grow becomes limited to $R_\mathrm{vlim}$ (Eq. \ref{eq:R_crit}) by either fragmentation or erosion. In this regime, the pebble opacity is instead given by:
\begin{subequations}
\begin{align}
    \kappa_\mathrm{vlim} &= \frac{3Q_\mathrm{eff} \rho_\mathrm{peb}}{4\rho_\mathrm{\bullet} R_\mathrm{vlim} \rho_\mathrm{g}}\\
    &= \left\{\begin{array}{l}
    \displaystyle\frac{3Q_\mathrm{eff}\dot{M}_\mathrm{peb} g}{16 \pi r^2 \rho_\mathrm{g}^2 v_\mathrm{th} v_\mathrm{lim}(v_\mathrm{lim}+v_\mathrm{gas})}
    \mathrm{\; (\mathrm{Epstein})}
    \\
    \\
   \displaystyle\frac{Q\dot{M}_\mathrm{peb}}{8\pi r^2 (v_\mathrm{lim}+v_\mathrm{gas})} \left(\frac{ g}{\rho_\mathrm{\bullet}\rho_\mathrm{g}^3 v_\mathrm{th} v_\mathrm{lim} l_\mathrm{mfp}}\right)^\frac{1}{2}
    \mathrm{\; (\mathrm{Stokes})}. \label{eq:opac_crit}
    \end{array}\right.
\end{align}
\end{subequations}
In combination, the pebble opacity can be approximated analytically from the expressions above by determining the appropriate regime based on the characteristic sizes, which have to be evaluated locally in the envelope:
\begin{equation}\label{eq:opac_peb}
    \kappa_\mathrm{peb} = \left\{\begin{array}{l}
    \kappa_\mathrm{vlim}
    \mathrm{\; if \;  } R_\mathrm{vlim} < R_\mathrm{coal}
    \\
    \\
   \kappa_\mathrm{coal}
    \mathrm{\; if \;  } R_\mathrm{vlim} > R_\mathrm{coal}.
    \end{array}\right.
\end{equation}
For completeness, we also include the simplest scenario where solids enter a planetary envelope with a constant size that remains unchanged during their sedimentation. While we follow the more physically motivated expressions from Eq. \ref{eq:opac_peb} in the rest of this work, the scenario of a constant pebble size might be applicable if pebbles are both sufficiently small to escape fragmentation and bounce rather than stick upon contact, as is the case if the pebbles are modeled as small molten chondrules, rather than dust agglomerates. Under this assumption, the opacity follows from Eq. \ref{eq:f_kappa}, \ref{eq:vsed} as (see also the work by \citet{Ormel2021}):
\begin{equation}\label{eq:opac_cst}
    \kappa_\mathrm{peb, cst} = \frac{3Q_\mathrm{eff}\dot{M}_\mathrm{peb}}{16 \pi r^2 \rho_\mathrm{\bullet} \rho_\mathrm{g} R_\mathrm{peb} v_\mathrm{sed, peb}}.
\end{equation}

\subsection{Steady state between dust replenishment and sweep-up.}\label{sect:equilibrium}
In previous grain growth models \citep{Ormel2014, Mordasini2014b}, a large influx of small dust grains in the envelope's outer layers was found to quickly diminish due to mutual sticking collisions. Hence, even when a significant fraction of the pebble mass is transferred to dust grains upon entry by efficient erosion (sect. \ref{sect:erosion}), the smallest particles soon disappear from the envelope, growing to the same limiting sizes as pebbles that we discussed in the previous section. The diminishing of the small grains abundance is further hastened by the sweep-up of pebbles below their erosion size. In order to maintain a dust population, therefore, it must be supplied by either continued erosion of pebbles that grow beyond the erosion limit, or by fragmentation. If that happens, the dust abundance has both a source and a sink term, generating a steady state when they are equal and opposite:
\begin{equation}
    \frac{d\dot{M}_\mathrm{d}}{dr}\bigg|_\mathrm{sweep-up} = -\frac{d\dot{M}_\mathrm{d}}{dr}\bigg|_\mathrm{frag},
\end{equation}
from which their respective volume densities follow as
\begin{equation}\label{eq:rho_d}
    \frac{\rho_\mathrm{d}}{\rho_\mathrm{peb}} = F x_\mathrm{R}.
\end{equation}
Eq. \ref{eq:rho_d} implies that in a steady state between fragmentation/erosion and sweep-up, most of the radial mass flux is generated by the larger sedimenting pebbles. But if $F$ is near unity, the volume density of grains can nevertheless be comparable to that of the pebbles due to their slower sedimentation.

\subsection{Dust opacity in steady state}\label{sect:opac_dust}
The dust opacity in steady state between dust production and sweep-up is given by:
\begin{subequations}
\begin{align}
    \kappa_\mathrm{d} &= \frac{3Q_\mathrm{eff, d} \rho_\mathrm{d}}{4\rho_\mathrm{\bullet} R_\mathrm{d} \rho_\mathrm{g}} \\
     &= \kappa_\mathrm{peb} \frac{Q_\mathrm{eff,d}}{Q_\mathrm{eff,peb}} \frac{R_\mathrm{peb}}{R_\mathrm{d}} F x_\mathrm{R}. \label{eq:opac_dust}
\end{align}
\end{subequations}
Eq. \ref{eq:opac_dust} is proportional to the pebble opacity, only differing from its trends due to the additional dependence on the pebble size. Because the opacity in planetary envelopes is generally far more variable than the pebble size, the opacity from dust in steady state generally follows a very similar trend to that of the pebbles. This makes it possible to evaluate the dust opacity, even though the parameter $F$ is largely unconstrained. In principle, it can vary between $0-1$, with a value around unity more appropriate in the erosion-limited regime where any pebble growth beyond the erosion limit is converted into dust. Lower values where collisions convert a smaller fraction of their mass into dust are likely more appropriate in cases where either fragmentation or growth (both pebble-pebble interactions) limits the pebble size. We will see in the next section that with our default value of $F=0.1$, the dust opacity is typically comparable to the pebble contribution. The biggest positive change with respect to the default model, therefore, occurs for $F=1$. For lower F, the opacity is not affected much because the pebble opacity then dominates and remains unchanged.

\section{Envelope opacity trends}\label{sect:opacity_trends}
\begin{figure}[t!] 
\centering
\includegraphics[width=\hsize]{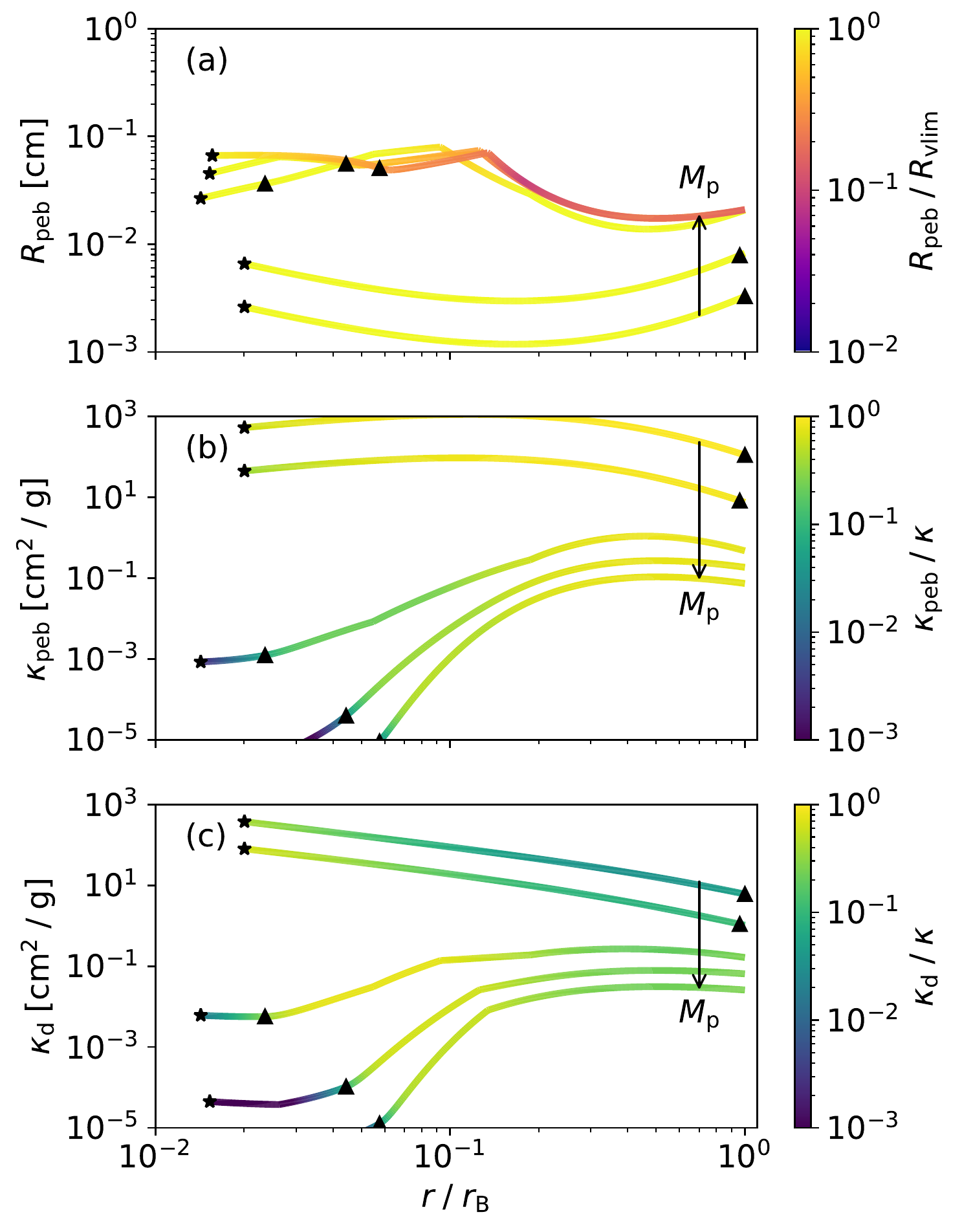}
\caption{Pebble growth tracks (a), their resulting pebble opacity (b) and produced dust opacity (c) for a standard set of model runs at 5 AU (see Table \ref{table_parameters}). The different lines indicate a range of planet masses, with the arrow indicating a logarithmic progression from $0.5-20 \; \mathrm{M_\oplus}$. The triangles indicate the location of the RCB, while the stars indicate the depth where the ambient temperature exceeds the sublimation temperature (2500 K) and the opacity from solids vanishes. The colors in the top panel show the ratio of the pebble size in the model relative to $R_\mathrm{vlim}$ (Eq. \ref{eq:R_crit}). The colors in the two lower panels show the relative value of the indicated opacity to the total opacity.}
\label{fig:M_tracks}
\end{figure}
\begin{figure}[t!] 
\centering
\includegraphics[width=\hsize]{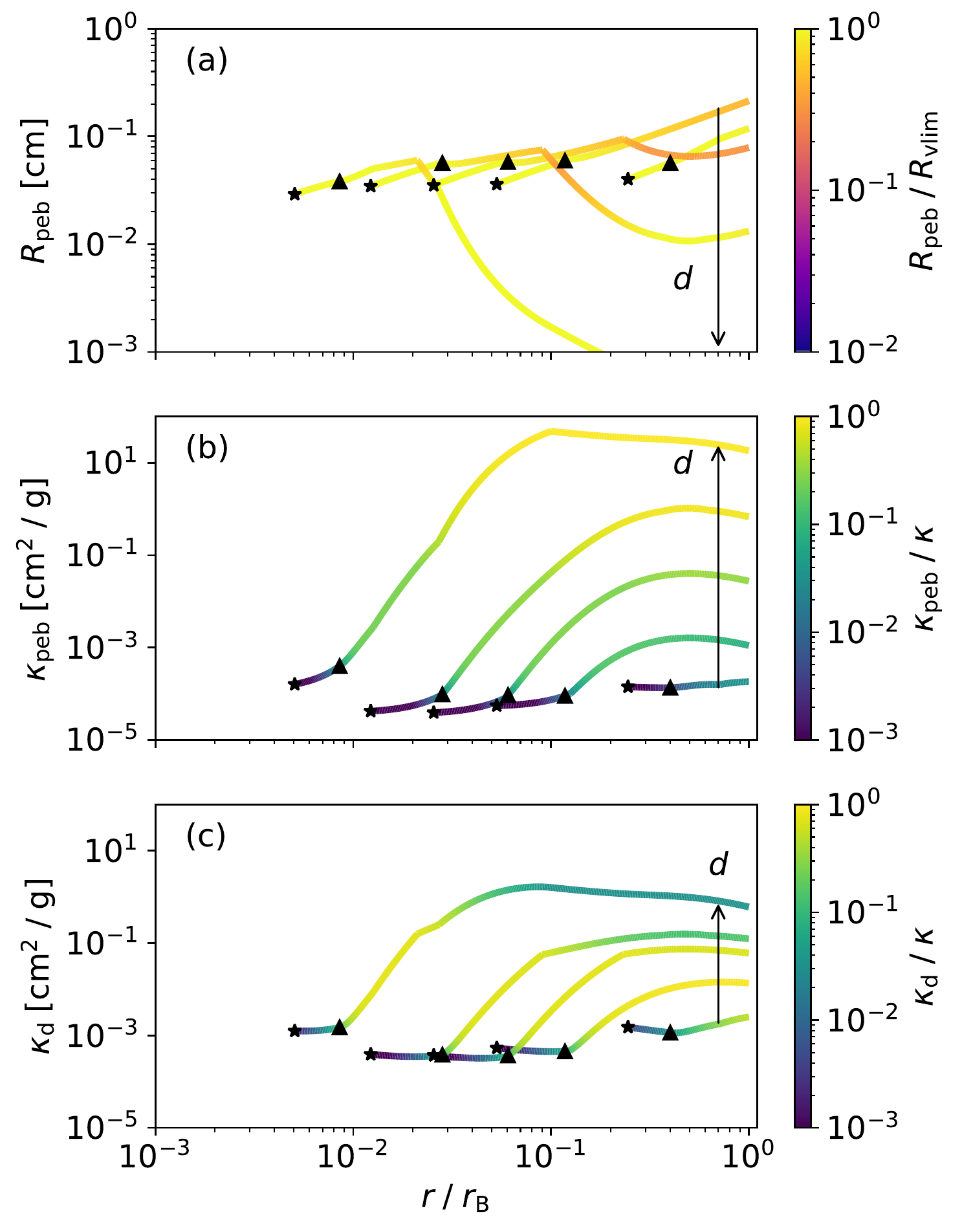}
\caption{Pebble growth tracks (a), their resulting pebble opacity (b) and produced dust opacity (c) for a standard set of model runs at 5 AU (see Table \ref{table_parameters}). This figure is the same as Fig. \ref{fig:M_tracks}, but now the mass is fixed at the default 5 $\mathrm{M_\oplus}$ and the planet's distances from the star are varied logarithmically from $0.1-30 \; \mathrm{AU}$.}
\label{fig:d_tracks}
\end{figure}
\begin{figure}[t!] 
\centering
\includegraphics[width=\hsize]{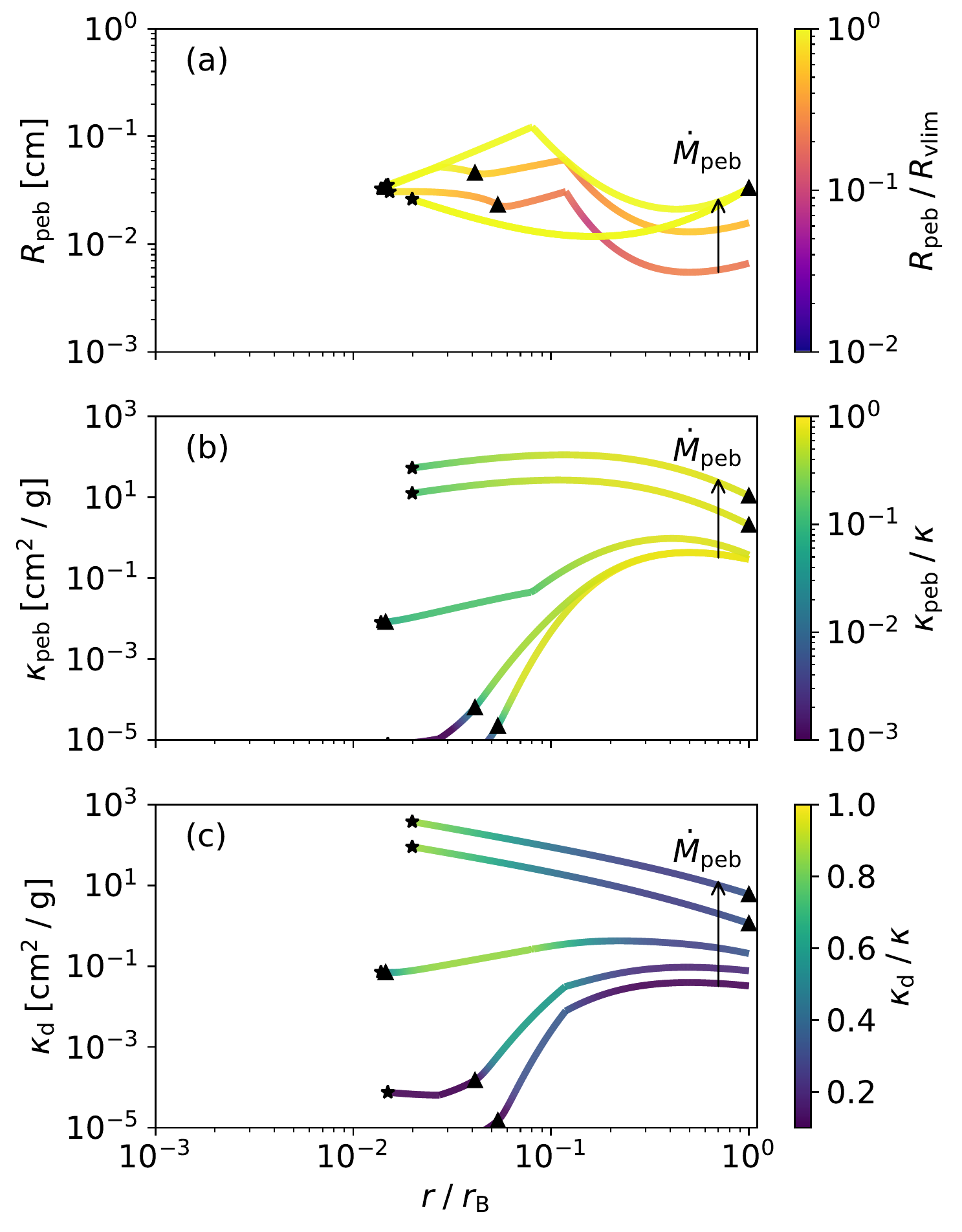}
\caption{Pebble growth tracks (a), their resulting pebble opacity (b) and produced dust opacity (c) for a standard set of model runs at 5 AU (see Table \ref{table_parameters}). This figure is the same as Fig. \ref{fig:M_tracks}, but now the mass is fixed at the default 5 $\mathrm{M_\oplus}$ and the pebble accretion rates are varied logarithmically from $10^{-7}-10^{-4} \; \mathrm{M_\oplus/yr}$.}
\label{fig:Mdot_tracks}
\end{figure}
\begin{figure*}[t!] 
\centering
\includegraphics[width=\hsize]{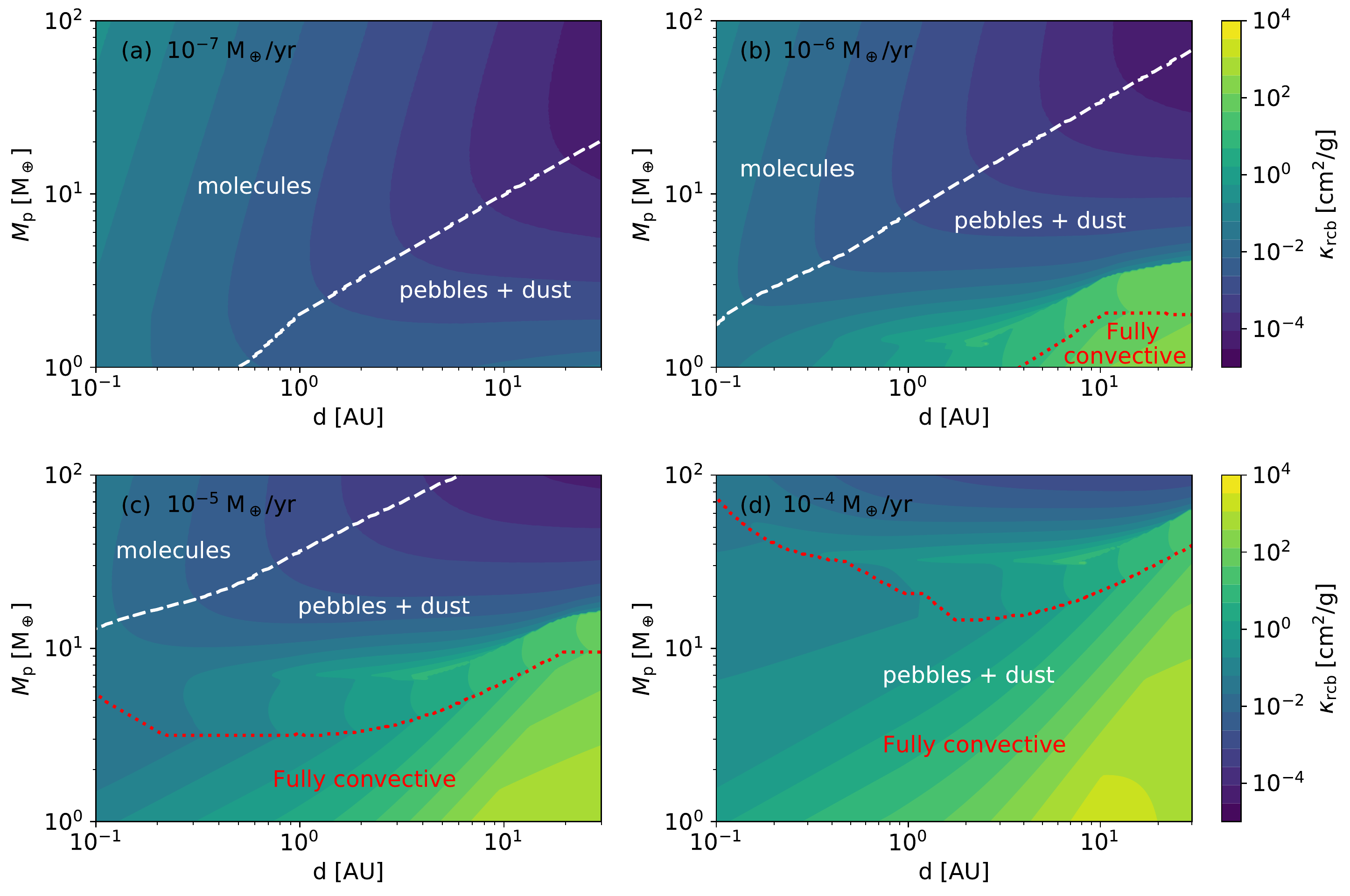} 
\caption{Compilation grid of $10^6$ runs whose colors indicate the opacity at the RCB as a function of the planet's semi-major axis and mass at four different pebble accretion rates. The white dashed lines divide the zones where different opacity contributions dominate. The red dotted lines mark the parameter space where the entire envelope is convective.}\label{fig:opacity_regimes}
\end{figure*}
In this section, we apply our model for the opacity of solids to a broad parameter space in order to investigate the its trends in envelopes of planets throughout the disk. We first detail our envelope model and then evaluate the opacity of dust and pebbles as a function of depth (Figs. \ref{fig:M_tracks}-\ref{fig:Mdot_tracks}) for a range of planetary masses, distances and pebble accretion rates. After that, we consider the opacity at the boundary between radiative and convective zones to visualize the same trends in a single graph (Fig. \ref{fig:opacity_regimes}).

\subsection{Envelope structure}\label{sect:envelope}
We focus our modeling efforts on the outer envelope down to the polluted region, ulterior to which no significant sublimation occurs and the opacity of solids is relevant. We refer to an accompanying paper by \citet{Ormel2021} and independent studies by \citet{Bodenheimer2018, Valletta2020} for detailed numerical models of polluted envelope interiors. The structure of the outer envelope is simple by contrast, as it shares its gaseous composition with the surrounding disk and is unaffected by self-gravity prior to the onset of runaway accretion. In quasi-hydrostatic equilibrium, its structure equations read:
\begin{subequations}
\begin{align}
    \frac{\partial m}{\partial r} &= 4\pi r^2 \rho_\mathrm{g}, \label{eq:dm_dr}\\
    \frac{\partial P_\mathrm{g}}{\partial r} &= -\frac{G M_\mathrm{p} \rho_\mathrm{g}}{r^2}, \label{eq:dP_dr} \\
    \frac{\partial T_\mathrm{g}}{\partial r} &= \frac{\partial P_\mathrm{g}}{\partial r} \frac{T_\mathrm{g}}{P_\mathrm{g}} \; \mathrm{min}\big(\nabla_\mathrm{rad}, \nabla_\mathrm{ad}\big), \label{eq:dT_dr}
\end{align}
\end{subequations}
where
\begin{equation}\label{eq:nabla_rad}
    \nabla_\mathrm{rad} = \frac{3 \kappa L P_\mathrm{g}}{64 \pi \bar{\sigma} G M_\mathrm{p} T_\mathrm{g}^4}
\end{equation}
is the radiative temperature gradient, which contains the gravitational and Stefan-Boltzmann constants ($G, \bar{\sigma}$) and is a function of the luminosity $L$ and the total Rosseland mean opacity $\kappa$. In our model, we assume a constant (global) accretion luminosity equal to $L=G M_\mathrm{c} \dot{M}_\mathrm{peb}/r_\mathrm{c}$ as is commonly done. However, we note the caveat that issued by \citet{Ormel2021} that this term can in reality vary substantially, as refractory material can be absorbed before sinking to the core and a significant portion of the gravitational energy can be processed in envelope mixing or used to heat the surrounding gas. 

The total envelope opacity is defined as the sum of the gas, dust and pebble contributions ($\kappa = \kappa_\mathrm{gas} + \kappa_\mathrm{d} + \kappa_\mathrm{peb}$). The gas opacity is often taken from lookup tables \citep[i.e.][]{Freedman2008, Freedman2014} that contain contributions from different gaseous species. We not not model this contribution here but choose to take a simple reference value of $\kappa_\mathrm{gas}$ as an analytical scaling of the molecular opacity from \citet{Bell1994}:
\begin{equation}
    \kappa_\mathrm{mol} = 10^{-8} \rho_\mathrm{g}^\frac{2}{3} T_\mathrm{g}^3 \; \mathrm{cm^2 g^{-1}},
\end{equation}
which was also used in the first two papers in this series. A simple molecular opacity scaling has the advantage that trends in the opacity from solids, which we seek to characterize here, can be more easily isolated in the results.

We integrate the envelope structure equations \ref{eq:dm_dr}-\ref{eq:dT_dr} from the planet's outer edge, which is taken as the minimum between the Hill and Bondi radii. The outer boundary conditions are equal to the local disk environment, for which we assume the simple temperature and density relations from the Minimum Mass Solar Nebula (see Table \ref{table_parameters}) \citep{Weidenschilling1977b, Hayashi1981}. The total opacity is explicitly included in the integration of the structure equations, and no iteration of the model is needed. We use the ideal gas equation to relate the local density to the pressure and temperature, which is justified in the upper layers of the atmosphere that are most important for the opacity calculation.

\subsection{Trend with planet mass}
In Fig. \ref{fig:M_tracks}, we plot the pebble sizes (panel a) along with their opacity (b) and the dust opacity (c) of our steady state model for a range of masses that we vary logarithmically between 0.5-20 $\mathrm{M_\oplus}$. The other parameters in the runs share are the default values from Table \ref{table_parameters}. We first discuss the top panel, which shows two separate trends depending on the process that limits the pebble size. In the velocity-limited regime ($R_\mathrm{peb}=R_\mathrm{vlim}$, yellow color range), the pebble size is limited by either the erosion with small dust grains or by collisions with other pebbles. Because the Bondi radius is positioned closer to the core in low-mass planets, pebbles of the same size are able to sediment faster ($g \propto M_\mathrm{p}^{-1}$) and the same terminal velocity limit leads larger pebbles in more massive planets. Once the planet becomes massive enough that $R_\mathrm{vlim}>R_\mathrm{coal}$, the pebble size ceases to be velocity-limited and instead becomes only limited by the rate of growth. In this growth regime, the pebble size is nearly invariant with a further increase of the planet mass (see Sect. \ref{sect:growth}).

Regardless of the regime, we find that the opacity from solids steeply declines when planets grow more massive. This can be explained with two competing processes. As planets accrete material, the surface area of their Bondi sphere increases quadratically with mass and the same pebbles obscure a smaller fraction of the envelope. Part of this trend is compensated by the slower sedimentation of pebbles in more massive planets but the net result is that the pebble volume density scales as $\rho_\mathrm{peb} \propto M_\mathrm{p}^{-1}$ at the Bondi radius. If the pebble size is instead limited by erosion and fragmentation, in which regime their sizes also scale positively with planet mass, the opacity from solids is an even steeper declining function. Note that the pebble and dust opacity trends are almost identical because they are nearly proportional to one another (Eq. \ref{eq:opac_dust}).

The difference in opacity between the low- and high-mass planets begins at around three orders of magnitude at the Bondi radius and these differences increase further with depth. Most of this increase is due to the transition from Epstein to Stokes drag that occurs in the more massive envelopes where larger pebbles enter a denser medium. With this, the difference at the Bondi radius is extended further, to over six orders of magnitude between the plotted values ($0.5-20 \; \mathrm{M_\oplus}$) at the radiative-convective boundary (RCB).

\subsection{Trend with orbital separation}
Next, we evaluate the importance of the planet's distance to the central star in Fig. \ref{fig:d_tracks}, where we plot five lines that indicate semi-major axes between 0.1-30 AU. The distance to the central star determines the local thermodynamic conditions of the surrounding disk. In our model, we use the Minimum Mass Solar Nebula, where the temperature and density scale as $T_\mathrm{disk} \propto d^{-1/2}, \; \rho_\mathrm{disk} \propto d^{-11/4}$ \citep{Weidenschilling1977b, Hayashi1981}. While the local conditions will also generally differ between different disk types and models, the general trend of lower densities and temperatures in the outer regions that receive less light is universal. This trend is important, because a more tenuous medium allows pebbles of the same size to sediment faster. The velocity-limited sizes, therefore, decrease with distance from the star and pebbles experience fewer collisions, reducing their rate of growth as well. This trend is reflected in Fig. \ref{fig:d_tracks}, where the opacity of solids at the Bondi radius is seen to increase with orbital separation, regardless of the regime.

The opacity trend with distance becomes more complex at greater depth, where the opacity of solids at the RCB is seen to converge to similar values for the parameters in Fig. \ref{fig:d_tracks}. In these runs, the pebble sizes of planets in the outer disk increase with depth as the surrounding medium becomes more dense and the same velocity limit allows for larger pebbles. As a result, the opacities also decline with depth until they are surpassed by the molecular contribution near the RCB, where the local pebble sizes become limited by growth with the parameters plotted here. For smaller planets in the outer disk, or those that are accreting pebbles at a higher rate, the pebble sizes at the RCB instead remain limited by fragmentation and erosion. We will look into the distance trend of this velocity-limited opacity regime in Sect. \ref{sect:dichotomy}.

\subsection{Trend with pebble accretion rate}\label{sect:mdot_scaling}
Finally, we examine the opacity trend with the pebble accretion rate in Fig. \ref{fig:Mdot_tracks}, where it is varied between $10^{-7}-10^{-4} \; \mathrm{M_\oplus/yr}$. This probes both the variation in solid mass flux and accretion luminosity, which are proportional ($L=G M_\mathrm{c} \dot{M}_\mathrm{peb}/r_\mathrm{c}$) in our model. The effect on the pebble sizes (top panel) can be divided into two regimes, depending on whether the pebble size is limited by growth (orange color range) or velocity (yellow color range). In the former case, the pebble size scales positively with the accretion rate as an increased pebble mass flux increases the number of collisions during their sedimentation and allows for faster growth. In this regime, the pebble opacity (per Eq. \ref{eq:opac_coal}) is not explicitly dependent on the mass flux, aside from the extinction efficiency which is higher for larger pebbles in these conditions. The second effect is that a higher luminosity increases the local radiative temperature gradient, which alters the conditions of the local medium, ultimately affecting the opacity as well.

When the accretion rate is increased beyond $\sim 10^{-5} \; \mathrm{M_\oplus/yr}$ for the plotted default parameters, the pebble size becomes velocity-limited at the fragmentation-erosion barrier. At this point, any additional material no longer increases the pebble size and directly increases their volume density, allowing the opacity to rise sharply. The increased opacity in these runs with high pebble accretion rates turns the envelopes almost entirely convective (see upper two curves), with RCB locations at or close to the Bondi radius.

\subsection{Description of three opacity regimes}\label{sect:dichotomy}
In order to visualize the broad opacity trends more clearly, we also evaluate a large 2D grid of envelope opacities at the RCB ($\kappa_\mathrm{rcb}$) as a function of their distance to the star and planetary mass. While the opacity throughout the entire radiative zone is relevant for the envelope's structure, its value at the boundary between the radiative and convective zones is generally considered the most important. This is both because most of the radiative portion of the envelope's mass is contained near the RCB and because it controls the rate at which the convective interior cools \citep[i.e.][]{Lee2015, Ginzburg2016}.

The results of these runs are shown in Fig. \ref{fig:opacity_regimes}, ordered into four panels that correspond to different pebble accretion rates, which typically increase as planets grow and capture pebbles more efficiently \citep{Ormel2010, Lambrechts2012, Liu2019}. The opacity trends reflect the discussions of the previous subsections but also contrast the opacity of solids and molecules, which generally follow a dichotomy based on the planet's distance and mass. Hot molecules contribute most of the envelope's opacity in the warm and dense inner disk, where solids sediment slowly and coalesce to form larger agglomerates. Because the molecular opacity increases with both temperature and density, this provides a clear contrast based on orbital separation. The molecular opacity at the RCB is not very sensitive to planet mass, as can be seen from the nearly vertical opacity contours. In contrast, the opacity from solids declines steeply as planets grow more massive and particles sediment more slowly due to the higher densities. Fig. \ref{fig:opacity_regimes} shows that the total envelope opacity can be divided into three regimes:
\begin{enumerate}
    \item The molecular opacity regime applies to higher-mass planets in the inner disk that are accreting pebbles at a low rate. In this regime, the opacity at the RCB is almost independent of planet mass and steeply declines with orbital separation.
    \item The growth-limited solids opacity regime applies to higher-mass planets in the outer disk that are accreting pebbles at a low rate. In this regime, the RCB opacity is almost independent with distance and declines steeply with planet mass. The opacity is invariant to an increase of the mass flux, as any additional mass just adds to pebble growth. However, the RCB opacity scales positively with the accretion luminosity.
    \item The velocity-limited solids opacity regime applies to lower-mass planets that are accreting pebbles at a sufficiently high rate. It is typically characterized by fully convective envelopes, with the plotted RCB located at the Bondi radius. A higher pebble accretion rate increases the volume density of solids in this regime and greatly increases the parameter range where it applies (to include larger $M_\mathrm{p}$ and smaller $d$). 
\end{enumerate}
One caveat to this third regime is that if the entire outer envelope is convective, the envelope density remains relatively low and as pointed out by \citet{Johansen2020}, the envelope can then locally become radiative at the depth where solids sublimate (2500 K in our model). At this point, which roughly coincides with the dissociation temperature of molecular hydrogen, only the molecular contribution to the opacity remains. If there are sufficient free electrons to ionize the hydrogen and produce $\mathrm{H}^{-}$, this becomes the dominant opacity component \citep{Lee2015}. The situation in this regime is further complicated by the balancing effect of a compositional gradient that also begins to form at these temperatures \citep{Bodenheimer2018, Muller2020}, in which the Ledoux criterion \citep{Ledoux1947} has to be used to evaluate stability against convection, rather than the Schwarzschild criterion. The thermodynamic consequences of this transition region are an active topic of investigation \citep{Valletta2020, Ormel2021} and are outside the scope of this opacity study, where we focus specifically on the opacity of solids in the outer envelope. We provide a broader discussion of this potential zone in Sect. \ref{sect:discussion}.

\section{Implications for giant planet formation}\label{sect:implications}
Planets transition from slow to runaway gas accretion when they reach a critical point where their self-gravity begins to exceed the envelope's pressure support. A transparent envelope is characterized by more efficient radiative energy transport in the outer regions, which results in faster cooling and more rapid gas accretion. As a result, both analytical and numerical works \citep[e.g.,][]{Mizuno1980, Stevenson1982, Pollack1996, Movshovitz2010, Lee2015} have shown that the mass at which planets transition to runaway growth is positively linked to the envelope's opacity. In this section, we apply our model to examine this trend across the proto-planetary disk.

\subsection{Trends in the critical metal mass}
\begin{figure}[t!] 
\centering
\includegraphics[width=\hsize]{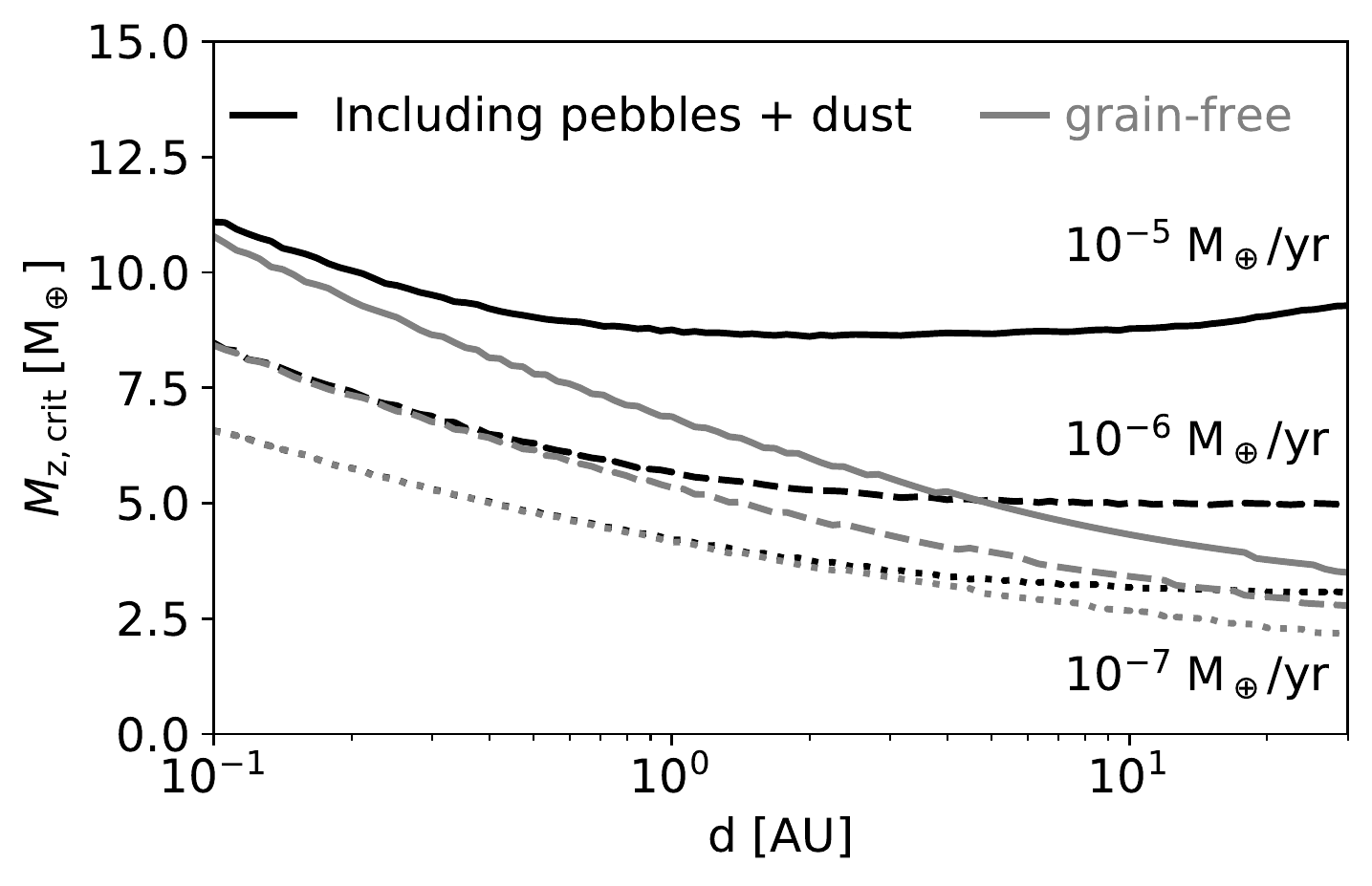} 
\caption{Trends in the critical metal mass as a function of distance. The different lines correspond to a variation in solids accretion rates: $10^{-7} \; \mathrm{M_\oplus/yr \; (dotted)}$, $10^{-6} \; \mathrm{M_\oplus/yr \; (dashed)}$, $10^{-5} \; \mathrm{M_\oplus/yr \; (solid}$), with proportional gas accretion $\dot{M}_\mathrm{xy}/\dot{M}_\mathrm{peb}=5$ and the remaining parameters set by Table \ref{table_parameters}. Planets with grain-free envelopes (gray lines) exhibit a downward trend with orbital separation because the molecular opacity scales positively with temperature. The opacity generated by pebbles and dust (black lines) increases the critical mass significantly in the outer disk, especially at higher accretion rates. \label{fig:M_crits}} 
\end{figure}

In a previous work \citep{Brouwers2020}, we showed that the traditional criterion of a critical core mass becomes meaningless if the growth of the core is limited. For this reason, we introduced and derived an analogous criterion called the \textit{critical metal mass}, which measures the total metal content of a planet at the onset of runaway gas accretion. It shares the opacity dependence with previous expressions for the critical core mass, but is additionally an explicit function of the core mass itself. One limitation of our previous work was that we modeled envelopes with an isothermal radiative layer, which is a good assumption for grain-free envelopes but invalid for envelopes with higher opacities due to the presence of solids. In order to correctly account for thermodynamic changes in the radiative part of the envelope, we slightly modify the analytical structure model of \citet{Brouwers2020} in Appendix \ref{appendix_A} to be applicable to envelopes with non-isothermal radiative regions. The critical metal mass, defined as the total mass of solids accreted at the onset of runaway accretion, then becomes an explicit function of both $T_\mathrm{rcb}$ and $\kappa_\mathrm{rcb}$:
\begin{align}
    M_\mathrm{z,crit} &\approx 5.5 \; \mathrm{M_\oplus} \; 
    {\left(\frac{\kappa_\mathrm{rcb}}{0.01 \; \mathrm{g \; cm^{-2}}}\right)}^\frac{1}{6}
    {\left(\frac{d}{\mathrm{AU}}\right)}^\frac{7}{108}
    {\left(\frac{T_\mathrm{vap}}{2500 \; \mathrm{K}}\right)}^\frac{8}{27}     \label{eq:M_crit_analytical_3}
     \\ 
    & \qquad \qquad \;\;
    {\left(\frac{\dot{M}_\mathrm{peb}}{10^{-6} \; \mathrm{M_\oplus} \; \mathrm{yr}^{-1}}\right)}^\frac{1}{6}
    {\left(\frac{M_\mathrm{c}}{\mathrm{M_\oplus}}\right)}^\frac{1}{2} {\left(\frac{T_\mathrm{rcb}}{T_\mathrm{disk}}\right)}^{-\frac{126}{972}}.
     \nonumber
\end{align}
In line with the findings by \citet{Ormel2021}, we assume core growth to be uniformly limited to $2 \; \mathrm{M_\oplus}$, with the rest of the solids being absorbed in the deep envelope. We plot the resultant critical metal masses as a function of orbital separation for three accretion rates in Fig. \ref{fig:M_crits}. The 
black lines include the opacity from dust and pebbles, contrasted with the gray lines that assume entirely grain-free envelopes.

The general shape of the curves in Fig. \ref{fig:M_crits} reflects the opacity trends described in the previous section. Molecules dominate the opacity of envelopes in the hot inner disk, where the critical mass is seen to decrease with orbital separation. In the grain-free curves, this downward trend continues towards the outer disk where the critical mass reaches very low values of only a few $\mathrm{M_\oplus}$. Accounting for the opacity of pebbles and dust breaks this trend in the intermediate disk, where solids begin to dominate the opacity of the envelopes. This halts the decline of the critical mass, which then instead flattens towards the outer disk. Besides its variation with distance, the critical metal mass is also positively dependent on the solids accretion rate. This is both due to the increased luminosity and the increased opacity from solids, although the two are linked thermodynamically (see Sect. \ref{sect:mdot_scaling}). We note that in reality, a pebble-accreting planet would not experience a single pebble accretion rate during its evolution, but one that generally increases as it gains in mass and is able to capture pebbles more efficiently.

Physically, the trends in the critical metal mass translate to similar trends in the occurrence rates of gas giant planets. Data from the Kepler mission has provided good exoplanet abundance statistics up to about 1 AU in semi-major axis \citep{Borucki2010,Borucki2011,Batalha2013}, which show a general increase in the abundance of giant planets with distance in this range \citep{Howard2012, Dong2013, Santerne2016}. When combined with data from radial-velocity measurements \citep{Mayor2011}, it shows a break in the giant planet occurrence rate between 2-3 AU, with a declining power law beyond this value \citep{Fernandes2019}. While other factors are certainly at play in planet formation across the proto-planetary disk, these trends are generally well matched by the critical metal mass predicted by our opacity model.

\subsection{Implications for the formation of Uranus and Neptune}\label{sect:ice_giants}
Far out in the disk, one of the key theoretical challenges is to explain how Uranus and Neptune accreted their substantial masses of $\sim 15 \; \mathrm{M_\oplus}$ without entering runaway growth \citep{Helled2014, Helled2020}. In the model developed by \citet{Lambrechts2014}, it is suggested that the accretion of solids continued throughout the disk lifetime and that the heat from impactors kept the envelope stable during its evolution. However, the planet's cold birth environments and low molecular opacity means that their accretion luminosity is easily radiated away, causing grain-free envelopes to implode into gas giants before they have grown beyond a few Earth masses. In order to prevent runaway accretion, the planets must either form when most of the nebula is gone \citep{Lee2016, Ogihara2020}, or their envelopes must still be sufficiently opaque that their heat does not escape so quickly. The common approach to this problem is to assume that the envelopes contain a large ISM-like opacity from small dust grains. In line with previous works on coagulation \citep{Movshovitz2008, Movshovitz2010, Mordasini2014b, Ormel2014}, our results indicate that such a high opacity is not realistic when grain growth determines the pebble sizes. It is possible, however, for planetary envelopes in the outer disk to attain these high opacities if the pebble sizes are velocity-limited by fragmentation or erosion. As we show in Fig. \ref{fig:opacity_regimes}, this requires the planet to continuously accrete pebbles at a sufficiently high rate.

Alternatively, planets begin to carve partial gaps around their orbit when they grow beyond several Earth masses \citep{Paardekooper2004, Paardekooper2006}. When they continue to grow, they locally invert the pressure gradient in the disk and stop accreting pebbles, a criterion which is known as the pebble isolation mass \citep{Morbidelli2015, Morbidelli2012, Ataiee2018}. As shown by \citet{Bitsch2015, Bitsch2018}, it is important to account for the cooling of the Solar nebula and the reduced disk scale height over time. In these simulations, the pebble disk scale height is found to reduce sufficiently in the first few Myr that both Uranus and Neptune can reach their isolation mass and stop accreting pebbles. This introduces a new issue however, as ceasing the supply of accretion heat can again easily trigger cooling and rapid gas accretion. Unlike the close-in super-Earths and mini-Neptunes which can be prevented from accreting gas by entropy advection \citep{Ormel2015b, Cimerman2017, Kurokawa2018, AliDib2020, Moldenhauer2020}, the ice giants will rapidly accrete gas due to their Kelvin-Helmholtz contraction. The solution offered by \citet{Alibert2018, Guilera2020} is that the reduction in pebble accretion can be compensated by the accretion of planetesimals, which could naturally form at the surrounding pressure bump. In a similar vein, \citet{Lambrechts2017, Chen2020, Bitsch2021} reason that while mm-cm pebbles are easily blocked by a local pressure bump, small dust ($\leq 200 \; \mathrm{\mu m}$) is sufficiently coupled to the gas to pass through \citep{Pinilla2016, Bitsch2018, Weber2018, Haugbolle2019} and provide the required opacity to prevent rapid cooling. Our results indicate that this solution essentially faces the same problem as continuous pebble accretion: Regardless of whether the mass is supplied by pebbles or planetesimals, the opacity from solids is insufficient to prevent runaway growth without a high accretion rate. In fact, the required planetesimal accretion rate is even higher than the equivalence in pebbles because the abundance of small solids is insufficient to reach the velocity-limited regime and the envelope opacity thus remains low.

The implication is that regardless of whether the ice giants became isolated from the pebble flow, their formation requires the constant accretion of material at high rates if they formed in-situ. This introduces a timescale problem, as they must then have formed in the window where they had enough time to accrete their observed masses, but not too early such that they accreted too much material and became gas giants. Accurately estimating the duration of this formation window requires more detailed models the ice giant's interiors, which are currently not yet well constrained \citep{Helled2020, Vazan2020}. In addition, it is important to account for the cooling of the solar nebula over time. The cooling of the surrounding disk has the same general effect as increasing the planet's distance from the central star (reduced $T_\mathrm{disk}, \rho_\mathrm{disk}$), which we showed increases the envelope opacity contribution of solids. In a future work, we plan to quantitatively estimate the required accretion rate and formation window of these ice giants.

\section{Discussion}\label{sect:discussion}
\subsection{Comparison with contemporary works}
In this work, we focused on the growth and destruction of solids as they travel through planetary envelopes, ignoring their evolution prior to accretion. A contrasting approach was taken in a contemporary work by \citet{Bitsch2021}, who modeled the size distribution of solids in the disk similar to \citet{Savvidou2020}, with the assumption that their distribution and opacity in the disk mid-plane also apply to the envelope's interior. We present evidence that the opacities at the RCB are actually very different from the disk due to a typically much higher gas temperature and density, as well as the significant size evolution of solids in the envelope. Larger pebbles fragment or face erosion when they enter planetary envelopes, while the lower end of the accreted size distribution experiences significant growth during sedimentation. Effectively, we predict much smaller envelope opacities in the growth-limited regime than \citet{Bitsch2021}, although high opacities remain possible in the velocity-limited regime. Nevertheless, while we argue that the sizes of solids in the envelope are disconnected from their size distribution in the disk, modeling their size evolution in the disk remains important for physical estimates of their accretion rates and efficiency.

An alternative approach to modeling the envelope opacity is presented by \citet{Chachan2021}, which is based on the work by \citet{Lee2015}. The authors argue that the sublimation of dust leads to a second radiative zone at the sublimation front, with an inner boundary around 2500 K where dissociation of $\mathrm{H}_2$ provides a new opacity source in the form of $\mathrm{H}^{-}$. The appearance of this inner radiative zone is based on the assumption that the envelope's outer layers are convective due to a high dust opacity. The requirement of free elections to form $\mathrm{H}^{-}$ means that the opacity studied by \citet{Chachan2021} is proportional to the local metalicity of the gas, which the authors take equal to the local dust-to-gas ratio of the disk, yielding an opacity minimum at intermediate distances (1-10 AU). While this is an interesting result, we note that the interior region around 2500 K is highly complex due to the compositional gradient that forms at this temperature, which affects the criterion for convective stability \citep{Ledoux1947, Muller2020}. The precise value of the metallicity at 2500 K is, therefore, hard to constrain without a detailed interior model that accounts for the saturated vapor pressure of the sublimated solids, which determines the envelope absorption of metals and is highly variable with temperature and composition \citep{Ormel2021}. Besides the compositional complexity, the appearance of a second radiative region furthermore requires a high dust opacity in the outer envelope, which we show is significantly reduced when coalescence is accounted for. Nevertheless, the work by \citet{Chachan2021} hints at a more complex scenario for the fully convective envelopes that we encounter in the velocity-limited opacity regime. The details of a potential deeper RCB are likely crucial in any model where it appears. It is, therefore, warranted to engage in a detailed study that also accounts for the important effects in this temperature range, including the varying composition of the envelope.

\subsection{Pebble vs planetesimal accretion}\label{sect:planetesimals}
Our work also highlights the difference between the formation of planets with pebbles or planetesimals. Whereas pebbles naturally provide an abundance of small grains in the outer envelope, planetesimals are much larger (>1 km, typically several hundred km, \citep[i.e.][]{Li2019, Klahr2020, Rucska2021} and only produce smaller grains when they experience significant mass loss or a dynamical fracture, where the pressure on the front of the planetesimal exceeds its strength and the planetesimal bursts apart. For this to occur, the planetesimal must be moving at very high velocity through a region with sufficient density, a condition that is only fulfilled near a planet's central core and far below the RCB \citep{Podolak1988, Mordasini2015, Pinhas2016, Brouwers2018, Valletta2019}. Similarly, mass ablation through frictional heating is inversely effective with planetesimal mass and only becomes important for planetesimal-sized objects close to the planet's central core. Besides this, the localized nature of planetesimal impacts also makes it more difficult to effectively obscure an envelope, allowing unaffected parts to continue radiating away energy.

\subsection{Model caveats and improvements}
The main simplification of our opacity model is the assumption that the distribution of the solids in a planetary envelope can be modeled with a characteristic local pebble size, supplemented with a constant-size dust population. This assumption has been shown to provide good agreement with coagulation simulations \citep{Mordasini2014b} and we argue that it is also justified when the pebbles are limited by a velocity threshold and pile up at a single size. Modeling a more complex size distribution self-consistently during sedimentation would require a computationally intensive approach with a large number of size bins and experimental data on the redistribution function of grains produced by both erosion and collisions. Furthermore, grains of different compositions are characterized by different strengths and sublimation temperatures, meaning that the maximum grain size will in reality fluctuate depending on the temperature regime where different species can survive. Finally, we considered quasi-static envelopes and did not model any convective overshoot from the envelope's deeper regions, which can potentially return a portion of the particles to the surface \citep{Popovas2018, Popovas2019} after they experience collisional grind-down from the additional acceleration \citep{AliDib2020, Johansen2020}.

\section{Summary and conclusions}\label{sect:conclusions}
In the scenario of pebble accretion, planets grow by the subsequent accretion of solids (micron-sized dust + larger pebbles) and gas. The amount of gas that a planet is able to bind at a certain mass depends on its cooling rate. This, in turn, is set by the envelope's opacity. In this work, we designed an opacity model that incorporates the main physical processes that influence the evolution of solids in planetary envelopes. Our approach can be considered an extension to previous works by \citet{Mordasini2014b} and \citet{Ormel2014}, as we model two populations of solids instead of one and add the effects of pebble erosion, fragmentation and dust sweep-up by pebbles to a grain growth prescription. We formulate convenient analytical expressions for the pebble opacity in different regimes (Eqs. \ref{eq:opac_coal}-\ref{eq:opac_cst}) and we estimate the dust opacity in a steady-state between dust sweep-up and production (Eq. \ref{eq:opac_dust}). Finally, we apply our model across a wide parameter space to map out the resulting opacity trends as a function of distance, planet mass and accretion rate. Our main findings are that:
\begin{enumerate}
    \item The initial size distribution of solids that a planet accretes is not an accurate representation of their sizes within the envelope. At the upper end of the distribution, we corroborate the recent findings by \citet{AliDib2020,Johansen2020} that larger pebbles are prone to erosion by dust grains when they accrete onto planetary envelopes. This erosion occurs as a result of the additional gravitational acceleration by the planet and does not require violent convective gas motion. The size limit at entry varies from a few micron in envelopes of small planets in the outer disk to several cm for more massive planets that are located closer to the central star.
    
    \item Within the envelope, the pebble sizes are limited by their rate of growth or by the velocity at which they begin to erode or fragment. The growth-limited regime applies to larger planets that are accreting pebbles at a low rate. As found by \citet{Mordasini2014b, Ormel2014}, the opacity in this regime drops far below ISM values, especially at greater depth or in the envelopes of more massive planets. Any additional pebbles that a planet accretes in the growth-limited opacity regime only lead to larger pebbles and do not increase the opacity, other than through any added luminosity. Nevertheless, the contribution of solids typically still dominates over the molecules, with the exception of higher-mass planets in the inner disk (see Fig. \ref{fig:opacity_regimes}). Smaller planets at greater distance from the star or those that are accreting pebbles at a sufficiently high rate enter the velocity-limited regime, where pebble growth is restricted by erosion or fragmentation. In this regime, a higher pebble accretion rate further increases the pebble abundance and the opacity is able to remain high throughout the envelope, typically turning it entirely convective.
    
    \item One proposed solution for the fact that Uranus and Neptune accreted their substantial masses of $\sim 15 \; \mathrm{M_\oplus}$ without entering runaway gas accretion is that they had a high envelope opacity during formation. We find that the commonly made assumption of an ISM-like opacity from small dust grains is not realistic when grain growth is accounted for. High envelope opacities remain possible if the pebble accretion rate is sufficiently high and the pebble sizes become limited by fragmentation or erosion, rather than by their rate of growth. However, this introduces a timescale problem, as the planets must have formed at the right time to grow to their current sizes at a sufficiently rapid pace without becoming gas giants.
\end{enumerate}
%

\section*{Acknowledgements}
\tiny{Part of the work presented here is based on discussions conducted at the second ISSI ice giants Meetings in Bern, March 2020. MGB acknowledges the support of a Royal Society Studentship, RG 160509. AB is grateful to the Royal Society for a Dorothy Hodgkin Fellowship. We thank an anonymous referee for comments that helped to substantially improve the clarity of this work.}

\bibliographystyle{aa}
\bibliography{opacity}
%

\begin{appendix}
\section{Expression of the critical metal mass with a non-isothermal radiative layer}\label{appendix_A}
\begin{figure}[t!] 
\centering
\includegraphics[width=\hsize]{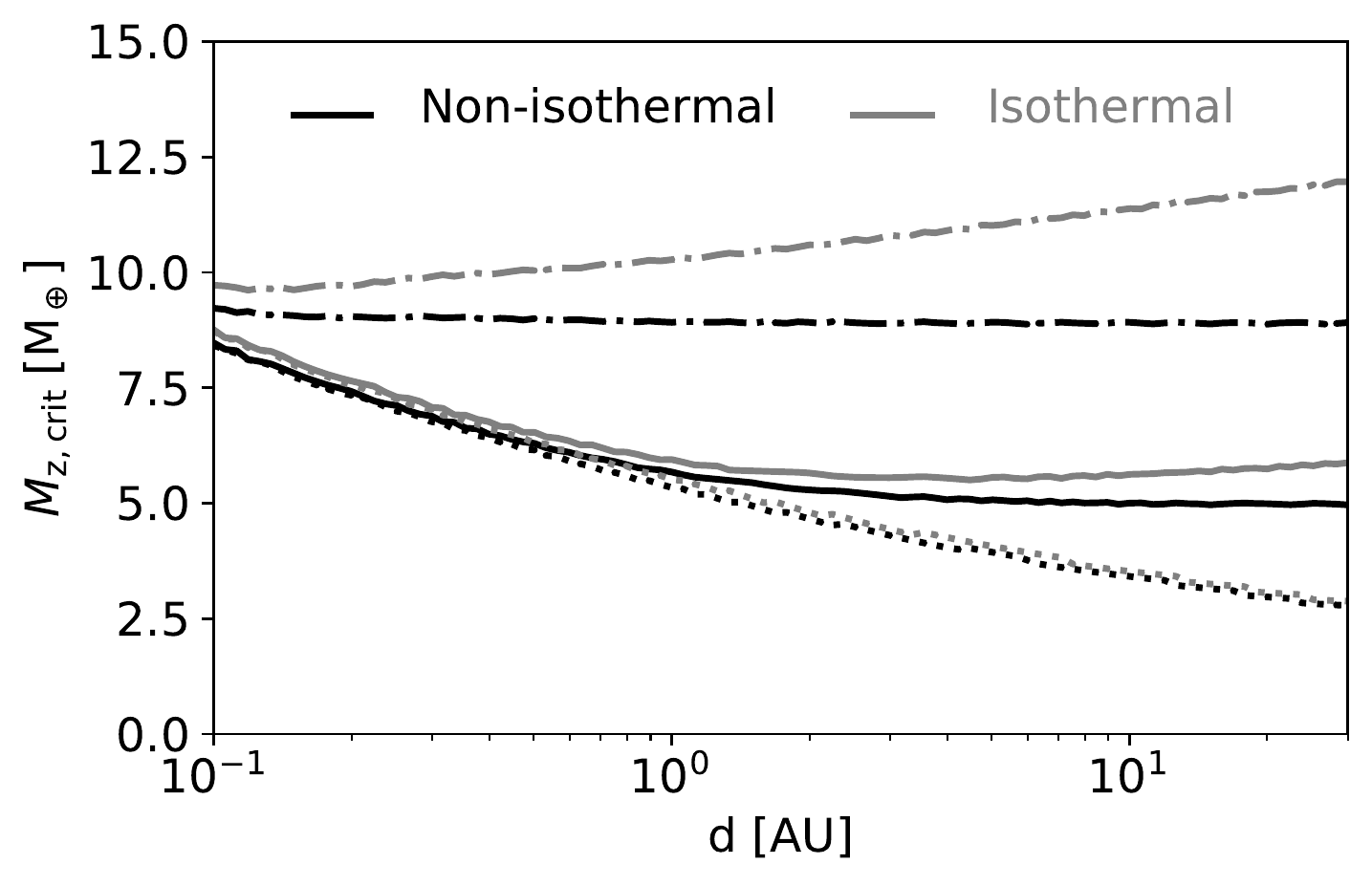} 
\caption{Comparison of the critical metal mass expressions of \citet{Brouwers2020} (isothermal, gray) and this work (non-isothermal, black), both examined with different opacities. The dotted lines indicate grain-free envelopes, whereas the solid lines assume the accretion of small pebbles that can grow, fragment and erode according to our model. The dash-dotted line corresponds to the accretion of $100 \; \mathrm{\mu m}$ pebbles that are kept at a constant size during their sedimentation. \label{fig:M_comparison}} 
\end{figure}

\normalsize{
In this appendix, we slightly modify the analytical structure model of \citet{Brouwers2020} to derive an expression for the critical metal mass that is more appropriate to study the effects of envelope opacity. The original model consists of an outer radiative region, followed by an intermediate convective region with solar composition and a polluted interior convective region that consists primarily of silicate vapor. Because that model was built with a grain-free envelope in mind, the radiative region was assumed to be isothermal. This ceases to be a good approximation when the opacity from solids is included and the radiative temperature gradient becomes important.

Fortunately, it is straightforward to modify the model to be applicable to a scenario with higher envelope opacities. We shift the temperature boundary of the intermediate convective region from $T_\mathrm{disk}$ to $T_\mathrm{rcb}$ and recalculate the thermodynamic structure of the intermediate layer. This amounts to a single variable change in Eqs. 9a-9c of \citet{Brouwers2020}:
\begin{equation}
    r'_\mathrm{B} = \frac{\gamma_\mathrm{xy}-1}{\gamma_\mathrm{xy}} r_\mathrm{B} 
    \quad \mathrm{to} \quad 
    r'_\mathrm{B} = \frac{\gamma_\mathrm{xy}-1}{\gamma_\mathrm{xy}} \frac{T_\mathrm{disk}}{T_\mathrm{rcb}} r_\mathrm{B} \quad
\end{equation}
which in effect lowers the density of the intermediate region and increases its temperature. The relation between the opacity and density at the RCB then becomes:
\begin{equation}
    \kappa_\mathrm{rcb} = \frac{64\pi\bar{\sigma} T_\mathrm{rcb}^4 r'_\mathrm{B}}{3 \rho_\mathrm{rcb} L}.
\end{equation}
The corresponding structure equations of the polluted region remain unchanged. The only modification we have to make is at the outer boundary of the polluted region, where the density changes from
\begin{equation}
     \rho_\mathrm{vap} = {\left(\frac{T_\mathrm{vap}}{T_\mathrm{disk}}\right)}^\frac{1}{\gamma_\mathrm{xy} - 1} \frac{\mu_\mathrm{g}}{\mu_\mathrm{xy}} \rho_\mathrm{rcb}
    \quad \mathrm{to} \quad 
    \rho_\mathrm{vap} = {\left(\frac{T_\mathrm{vap}}{T_\mathrm{rcb}}\right)}^\frac{1}{\gamma_\mathrm{xy} - 1} \frac{\mu_\mathrm{g}}{\mu_\mathrm{xy}} \rho_\mathrm{rcb}.
\end{equation}
In our model for a polluted envelope, most of the mass is contained in the inner regions. In order to derive the critical metal mass, we substitute the modified variables into Eq. 22b of \citet{Brouwers2020} and follow the same steps as in Sect. 4.2 of that work. In practice, this comes down to substituting $T_\mathrm{rcb}$ for $T_\mathrm{d}$ in their Eq. 27. The critical metal mass is approximated to occur at the crossover mass where $M_\mathrm{xy} = M_\mathrm{z}$, which yields:
\begin{align}
    M_\mathrm{z,crit} &\approx 5.5 \; \mathrm{M_\oplus} \; 
    {\left(\frac{\kappa_\mathrm{rcb}}{0.01 \; \mathrm{g \; cm^{-2}}}\right)}^\frac{1}{6}
    {\left(\frac{d}{\mathrm{AU}}\right)}^\frac{7}{108}
    {\left(\frac{T_\mathrm{vap}}{2500 \; \mathrm{K}}\right)}^\frac{8}{27} \label{eq:M_crit_analytical_appendix}
    \\ 
    & \qquad \qquad \;\;
    {\left(\frac{\dot{M}_\mathrm{peb}}{10^{-6} \; \mathrm{M_\oplus} \; \mathrm{yr}^{-1}}\right)}^\frac{1}{6}
    {\left(\frac{M_\mathrm{c}}{\mathrm{M_\oplus}}\right)}^\frac{1}{2} {\left(\frac{T_\mathrm{rcb}}{T_\mathrm{disk}}\right)}^{-\frac{126}{972}}.
     \nonumber
\end{align}
The only difference in the final expression is the appearance of an inverse scaling with $T_\mathrm{rcb}/T_\mathrm{disk}$. This is due to a combination of two opposing effects. The first effect is that a higher RCB temperature leads to a lower RCB density at the same pressure and, therefore, results in an increased critical mass. The second trend is that a higher RCB temperature moves the location of the RCB inward to a higher pressure and an increased density at the same opacity, leading to a reduced critical mass. In combination, these two trends yield a slight inverse scaling in the final expression. We show a comparison with the expression from \citet{Brouwers2020} in Fig. \ref{fig:M_comparison}, where the non-isothermal expression is shown in black as opposed to gray for the original expression. As expected, the difference is greatest when there is a large opacity from solids. Grain-free envelopes remain well approximated by the isothermal model. This is because their opacity scales positively with temperature and density, such that their radiative temperature gradients are steep functions of depth. We also plot the critical metal mass for a run where we keep the pebble size constant at 100 $\; \mathrm{\mu m}$. In agreement with the numerical results from \citet{Ormel2021}, we find that this yields a flat line. The quantitative values from Eq. \ref{eq:M_crit_analytical_appendix} are generally a factor $\sim$ 2 reduced compared to detailed numerical simulations with non-ideal equations of state performed by \citet{Ormel2021} but the simple derived scaling captures the main physical trends well.}

\section{Justification for neglecting dust growth in production-sweep-up steady state}\label{sect:appendix_B}
\normalsize{In this appendix, we investigate whether dust growth can be neglected in our steady-state model of dust production and sweep-up. To begin, we estimate the typical distance that dust grains travel before mutually colliding ($l_\mathrm{col,d}$), which is analogous to Eq. \ref{eq:l_col} for pebbles:}
\begin{equation}
    l_\mathrm{col,d} =  \frac{m_\mathrm{d}}{\rho_\mathrm{d}\sigma_\mathrm{d}} \frac{v_\mathrm{sed,d}}{v_\mathrm{col,d}}.
\end{equation}
\normalsize{
We compare this quantity with the typical distance that these dust grains can travel before they are swept up by a sedimenting pebble ($l_\mathrm{sweep,d}$), which in turn is analogous to Eq. \ref{eq:l_sweep} for pebbles:}
\begin{equation}
    l_\mathrm{sweep,d} = \frac{m_\mathrm{peb}}{\rho_\mathrm{peb}\sigma_\mathrm{peb}} \frac{v_\mathrm{sed,d}}{v_\mathrm{fall,peb}},
\end{equation}
\normalsize{
where we assumed that $v_\mathrm{fall,peb} \gg v_\mathrm{fall,d}$. As discussed in Sect. \ref{sect:equilibrium}, the dust and pebble volume densities in a steady state between dust sweep-up sweep-up and production are related through Eq. \ref{eq:rho_d}. This gives the ratio of the relevant length scales as:
\begin{align}
    \frac{l_\mathrm{col,d}}{l_\mathrm{sweep,d}} &= F^{-1} x_\mathrm{R}^{-2} \frac{R_\mathrm{d} v_\mathrm{fall,peb}}{R_\mathrm{peb} v_\mathrm{fall,d}} \\
    & \leq \frac{9}{F},
\end{align}
where the second line substitutes the scaling of Epstein drag, the drag regime with the most dust collisions. It is evident that even in this regime, $l_\mathrm{col,d}>l_\mathrm{sweep,d}$ and the dust grains are typically swept up by pebbles before they collide with other dust grains. This remains valid with the highest possible value of the dust production constant F = 1, so we conclude that significant dust growth is not expected in a steady state between dust production and sweep-up.

}
\section{Variation of the limiting velocity}\label{sect:appendix_C}
\begin{figure}[t!] 
\centering
\includegraphics[width=\hsize]{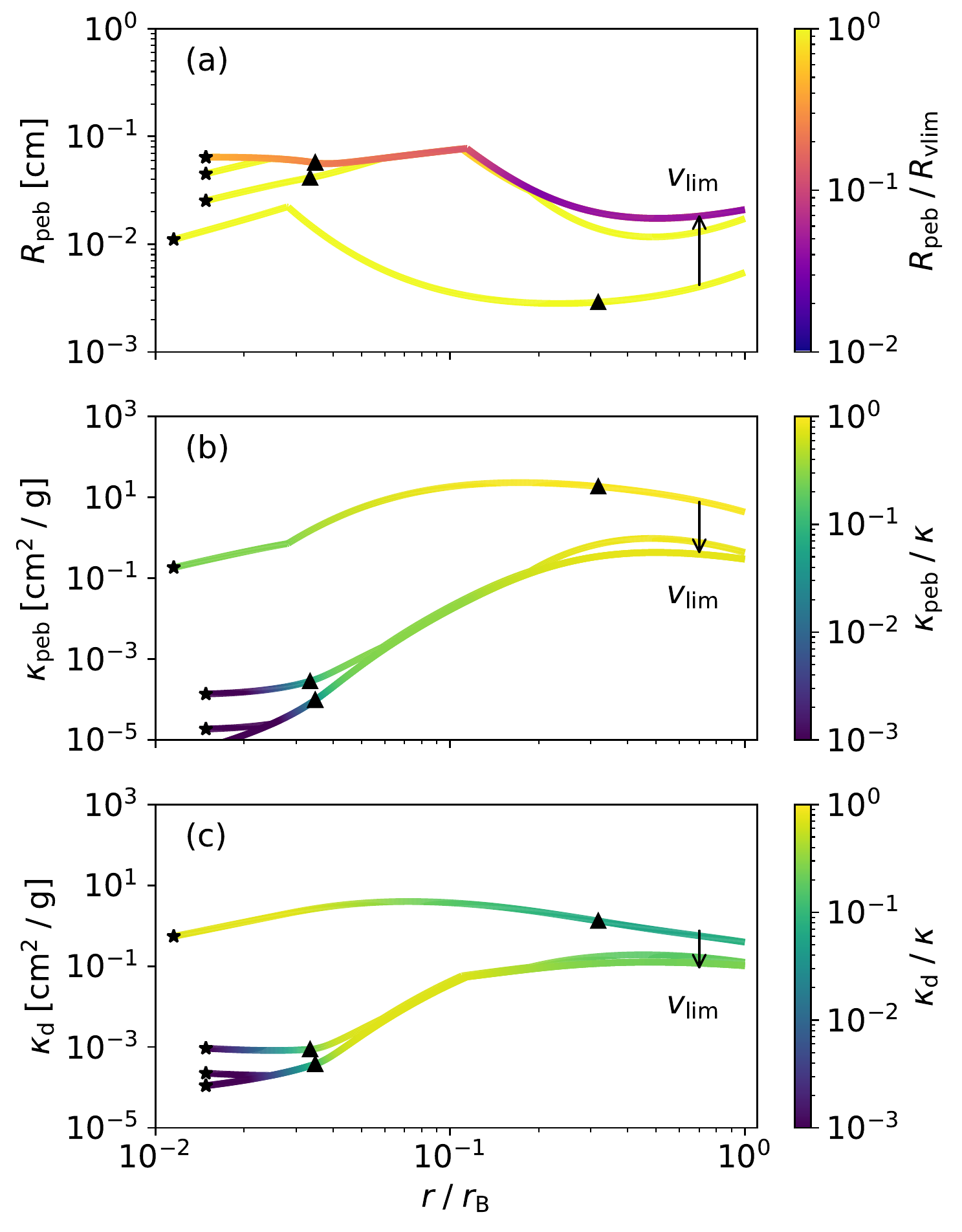}
\caption{Pebble growth tracks (a), their resulting pebble opacity (b) and produced dust opacity (c) for a standard set of model runs at 5 AU (see Table \ref{table_parameters}). This figure is the same as Fig. \ref{fig:M_tracks}, but now the mass is fixed at the default 5 $\mathrm{M_\oplus}$ and the limiting velocity (Eq. \ref{eq:vlim}) is varied a logarithmically between $0.4-40 \; \mathrm{m/s}$.}
\label{fig:v_tracks}
\end{figure}

\normalsize{The most important free parameter in our model is the limiting velocity (Eq. \ref{eq:vlim}), which represents the upper limit to the terminal velocity of pebbles during their sedimentation. It combines the velocity limits from erosion (Eq. \ref{eq:erosion_1}) and fragmentation. In the main text, we assumed a limiting velocity of $2.4$ m/s, corresponding to the onset of erosion by micron-sized dust grains \citep{Schrapler2018}. Because smaller dust grains are more effective at removing mass, the erosion velocity scales inversely with grain size. In fig. \ref{fig:v_tracks}, we vary the limiting velocity between 0.4-40 m/s, which corresponds to erosion from grain sizes between 0.05-100 $\mathrm{\mu m}$.

Fig. \ref{fig:v_tracks} illustrates that the value of the critical velocity has important consequences for the opacity. If the limiting velocity is reduced below our default value of 2.4 m/s, the pebbles are velocity-limited to smaller sizes and the opacity can significantly increase. Due to this, a broader range of envelopes become fully convective when they experience a given pebble accretion rate. The opposite trend is also true when the limiting velocity is increased. A higher velocity limit means that pebbles are more often limited by growth, which in turn means that higher accretion rates are required to keep envelopes of the same mass convective. Within an envelope whose sedimenting pebbles are already limited by growth, a further increase of the limiting velocity has no effect.}

\end{appendix}
\end{document}